# Structural effects of liquid infiltration of 3Y-Zirconia with Sc, Mg and Y

Asbjørn Slagtern Fjellvåg[1*], Øystein Slagtern Fjellvåg[2], Amund Ruud[1,3]

[1]Nordic Institute of Dental Materials (NIOM), Sognsveien 70 A, 0855 Oslo, Norway

[2]Department of Hydrogen Technology, Institute for Energy Technology (IFE), P.O. Box 40, NO-2027 Kjeller, Norway

[3]Department of Battery Technology, Institute for Energy Technology (IFE), P.O. Box 40, NO-2027 Kjeller, Norway

*corresponding author, a.s.fjellvag@niom.no

## 1. Abstract

The current work has investigated the effect of co-doping 3Y-Zirconia (3YSZ) with Sc, Mg and Y by liquid infiltration. Pre-sintered porous discs of 3YSZ were immersed in diluted nitric acid solutions containing Sc, Mg or Y, and combinations of the three, trapping liquid within the porosities of the samples. Upon drying, the cations are maintained inside the pellet, making the basis for the co-doping. After sintering, mass increase confirms the co-doping effect. Rietveld refinements of X-ray diffraction data show that the liquid-infiltrated samples contain the tetragonal *t*, *t''* and cubic *c*-phase in various fractions depending on the doping elements. The phases are compared with 3YSZ and 5YSZ sintered for 0 hours and 2 hours at 1500 °C. To evaluate the material properties in an application-based perspective, the Knoop hardness, translucency and grain size was measured. We conclude that liquid infiltration is a viable route to perform co-doping of Zirconia.

## 2. Introduction

The structural flexibility and applicable functional properties of Zirconia have received much attention for several decades. Different elements have been investigated for substitution of Zirconia, but co-substitution with multiple elements is less investigated due to the number of possible elemental combinations. Although it is tempting to map the structure-composition landscape of co-substituted Zirconia, the number of substituents and sample quality needed, with respect to *e.g.* particle size and successful processing, would be extremely time consuming. A simpler alternative for producing multiple different samples is liquid infiltration, utilizing already optimized Zirconia synthesis with small and uniform particles sizes. A micro-porous ceramic, such as pre-sintered Zirconia, is simply infiltrated with a precursor solution. The method was successfully used by Duh and Wan already in 1993 [1] for the co-substitution of Zirconia with $Y_2O_3$ and $CeO_2$ with the purpose of surface modification of the particles after synthesis. This approach is applicable for *e.g.* colouring of dental Zirconia by colour liquids [2–5], synthesis of various composites [6–11] and substituted ceramics [1,12–14], and for enhancing the translucency of ceramics [15,16].

Liquid infiltration has the advantage of high clinical relevance for dentistry, as it would be easy to apply a solution that enhances the translucency of dental restorations on a case-by-case basis. With that in mind, translucency effects are mainly published through several recent patents [17–20] that claim enhanced translucency with limited loss of strength by liquid infiltration with Y, Yb, Mg, Gd, La, Tm, or Ca. The solution can be applied using any method, such as brushing, dipping, ink-jetting, spraying, or specialized pens. Enhanced strength of 3-5Y-Zirconia by application of a Ta or Nb based solution is also patented [21], and the opposite effect of masking the translucent Zirconia [22].

The development of translucency enhancing liquids with significant effects is challenging. The optical properties of Zirconia trend with increasing Y-content in the sample, but are truly determined by the atomic structure, grain structure and porosity [23–25]. A deeper understanding of both the effect of co-substitution on Zirconia in general, and the effect of liquid infiltration on atomic- and grain structure is needed.

This paper investigates the effect of the substituent elements Mg, Y, and Sc on the atomic structure of co-substituted 3Y-Zirconia via liquid infiltration, targeting a total substituent concentration of about 10 mol. % in $ZrO_2$, like in 5YSZ. Upon infiltration of Mg, Sc and Y into 3YSZ we expect diffusion of the dopant into the YSZ lattice during sintering and that the cations occupy the Zr-position, forming oxygen vacancies as charge compensation (Table S4). We expect that a small cation like Sc(III) should cause higher tetragonality compared to Y(III), and that the higher number of oxygen vacancies from the lower valent Mg(II) should cause lower tetragonality than from Sc(III) [26–28].

The infiltrated samples were compared with commercial 3Y- and 5Y-Zirconia sintered for 2 hours (slow cooling) and 0 hours (quenched in air). All samples were analysed by synchrotron radiation X-ray diffraction (XRD) using Rietveld refinements for structural analysis, scanning electron microscopy (SEM) for qualitative grain size evaluation, Knoop hardness (HK1) for mechanical evaluation, and translucency parameter (TP) to evaluate the optical properties.

## 2.1. Background on atomic structure of Zirconia

$ZrO_2$ ($P2_1/c$) is a monoclinic and heavily distorted variant of the $CaF_2$-structure at room temperature, but transform to a tetragonal variant ($P4_2nmc$, Figure 1a) above ~1170 °C [29] and become cubic (Fm-3m, Figure 1b) above ~2370 °C [30]. Both the tetragonal and cubic phases can be stabilized at room temperature by substitution of several different oxides ($Y_2O_3$, MgO, CaO, $CeO_2$, $Al_2O_3$, etc.). The tetragonal stabilization is caused by disruption of the monoclinic distortions of the oxygen lattice, either by oxygen vacancies adjacent to Zr(IV) relieving the monoclinic distortions or by substituted cations associated with local distortions that relax neighbouring oxygen anions into 8-fold coordination around Zr(IV) [26–28,30–32].

Stabilization of the tetragonal phase by substitution in the range of 6 mol. % $YO_{1.5}$ or 10 mol. % $CeO_2$ is particularly beneficial for mechanical properties. The tensile stress at or near a crack tip can induce the transformation of the tetragonal ($t$) phase to monoclinic ($m$) structure ($t \rightarrow m$) [33]. The transformation gives rise to a 4-5 % volume expansion, which by the compressive stress applied from the surrounding grains reduces further crack propagation [34]. This transformation, known as transformation toughening, is the key to the use of Zirconia in applications requiring good mechanical properties and is reflected in the high strength and fracture toughness of different Zirconia grades [35,36] and composites [37,38]. This applies only as long as the $t$-phase does not spontaneously transform to $m$ at low temperature or in the presence of humidity, known as low temperature degradation (LTD) [30,39]. A small grain size can also stabilize the tetragonal phase at room temperature due to reduced surface energy compared to the monoclinic phase [34], enhancing LTD-resistance [40] and aiding mechanical properties by crack deflection [41,42]. Although a high transformability to $m$ results in high fracture toughness, it can come at the cost of flexural strength. This can, however, also be mitigated by a small grain size. Recently, 1.5Y-Zirconia (3 mol % $YO_{1.5}$) [35,40,43] and 5Ca-Zirconia (5 mol. % CaO) [44] have been reported with fracture toughness well beyond that of 3Y-Zirconia while maintaining high strength.

Two additional tetragonal structures (*t′* and *t″, $P4_2/nmc$*) are established as other versions of the tetragonal *t*-phase [45]. They are identical in symmetry but have different Y-content ($t < t' < t''$) and tetragonal distortion ($t > t' > t''$), see Figure 1. Investigations of 7-8 mol % $YO_{1.5}$ Zirconia by Krogstad *et al.* [32,45–48] concluded that during sintering, although having a homogeneous starting material, Y-diffusion towards the grain boundaries causes segregation of the *t′*-phase into Y-rich (*t″*) and Y-lean (*t*) regions. They also showed a linear correlation between tetragonality and Y-concentration, implying that *t*, *t′* and *t″* are not really different phases, merely the result of Y-diffusion driven by the thermodynamic two-phase region at the sintering temperature in the phase diagram. The main reason for considering these as different phases is that sufficient sintering time results in phase segregation into these phases.

The t′ and t″ phases have a lower tetragonality than $t$ (*t*~1.016, *t′*~1.011, *t″*~1.004 [46]) and are notably not transformable to *m* under tensile stress, unlike *t*. A sufficiently high tetragonal distortion is needed for the $t \rightarrow m$ transition to occur, related to the relative thermodynamic stabilities of *t* and *m* determined by the composition and grain size [37,49]. In commercial 3Y-Zirconia (3 mol. % $Y_2O_3$, 6 mol. % $YO_{1.5}$), approx. 80 % of *t* and 20 % of non-transformable phase is present, which could be *t′*, *t″* or *c* depending on the degree of phase partitioning [48,50]. Zirconia with more Y contains less *t* and more *t′*/*t″*/*c*, and has a corresponding reduced strength. The *t′*-phase is, however, associated with a ferro-elastic toughening mechanism in which the domains can switch under load, caused by a slight shift of the O-atoms. This provides enhanced mechanical properties, but not at the level of $t \rightarrow m$ transformation toughening. 7-8 mol % $YO_{1.5}$ Zirconia with *t′* is the preferred grade for thermal barrier coatings since it can withstand numerous temperature cycles with minimal to no structural degradation [30,31]. TEM investigations revealed that *t′* did not exist after the completed phase segregation for thermal barrier coating, but that it was an artifact present in the synchrotron radiation XRD data caused by the coherence of *t* and *t″*/*c* phases [48]. This is notably the same composition as the 4Y-Zirconia used in dental applications [51].

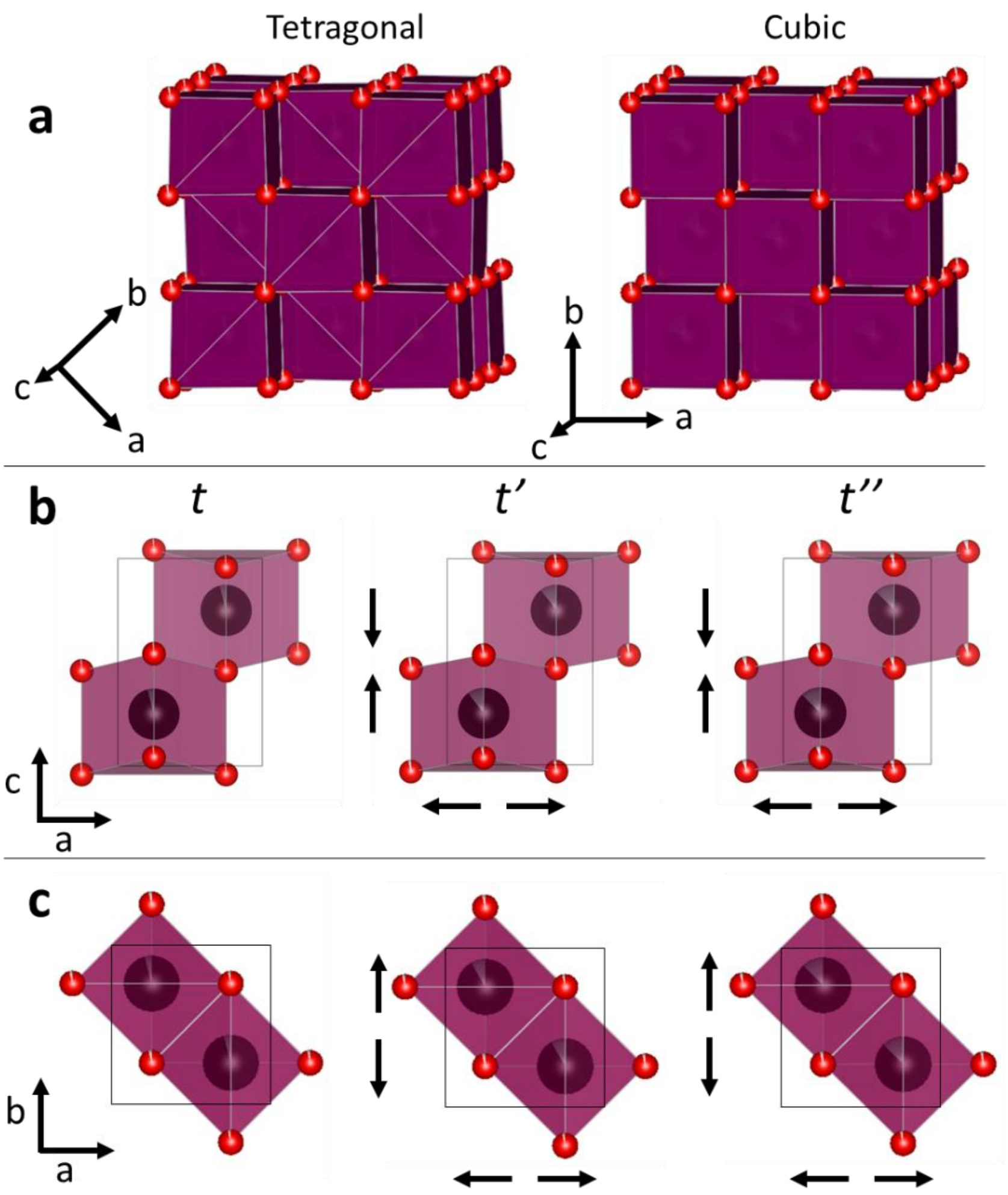

*Figure 1. Illustration of the Zirconia structure. a) overview showing the 8-coordinated cubes of the tetragonal and cubic structures, b) the t, t' and t'' phases seen along the b-axis and c) the t, t' and t'' phases seen along the c-axis. The arrows show the structural changes that occurs when moving from t to t' to t''. The illustrations are made using Vesta [52]. The purple circles represent Zr, the red circles represent O and the purple boxes represent the coordination polyhedron of O-atoms around Zr.*

# 3. Experimental

## 3.1. Samples

The starting material used for the infiltration experiments was commercial yttria-stabilized zirconia (YSZ) powder with 3 mol% $Y_2O_3$ in $ZrO_2$ (3YSZ) purchased from Tosoh (TZ-3YSB-E, Tosoh Europe B.V., Amsterdam, Netherlands). Small cylinders of approx. Ø20x20 mm were pressed uniaxially from approx. 20 g of YSZ in an in-house made cylindrical press (2.0 cm diameter) with 150 Mpa pressure, followed by cold isostatic pressing in water in an in-house made cylindrical press, at 200 MPa pressure. The pellets were pre-sintered with a standard heating sequence from room temperature to 1000 °C with a heating rate of 1 °C/min followed by a 2-hour dwell at 1000 °C and 10 °C/min cooling (Zirkonofen 700 Vakuum, Zirkonzahn GmbH, Austria). The pre-sintered pellets were thereafter sectioned into discs with thickness of approx. 0.6 mm on an automatic cutting device (Struers Secotom 60, Struers ApS, Ballerup, Denmark) with a diamond blade (M0D15 127 mm (5") dia. x 0.4 mm x 12.7 mm dia., Struers ApS, Ballerup, Denmark).

3YSZ (Tosoh TZ-3YSB-E) and 5YSZ (Tosoh Zpex Smile) reference samples were produced as described above. The samples were sintered at 1500 °C for 0- and 2-hours (Zirkonofen 700 Vakuum, Zirkonzahn GmbH, Austria), producing a total for four samples. The heating rate was 10 °C/min for both samples the cooling rate was 10 °C/min for the 2-hour samples and natural cooling (furnace switched off) for the 0-hour samples.

Magnesium oxide (MgO) (CAS: 1309-48-4, Product number from Sigma-Aldrich (PNSA): 243388-25G), Scandium(III)oxide ($Sc_2O_3$) (CAS: 12060-08-1, PNSA: 307874-5G), and Yttrium(III)oxide ($Y_2O_3$) (CAS: 1314-36-9, PNSA: 205168-50G), were all purchased from Sigma-Aldrich (Merck KGaA, Darmstadt, Germany). All oxides were dried at 1000 °C and cooled in a desiccator before weighing. Nitrate solutions, or Mg-, Sc-, and Y-solutions, respectively, were prepared by dissolving MgO and $Y_2O_3$ in diluted nitric acid at room temperature, and $Sc_2O_3$ in a large surplus of concentrated nitric acid at boiling temperatures. The Sc solution was further boiled to increase the Sc concentration in the solution through evaporation. Based on the mass of oxide dissolved of each element, solutions of 50:50 mol. % of Mg:Y, Mg:Sc and Y:Sc were made, with the same total concentration as the separate solutions of Mg, Sc and Y. The same applies to a mixed solution with 33.3 mol. % of each element (Mg:Y:Sc).

Several rounds of infiltration were performed. The infiltration was optimized towards a total dopant concentration of about 5 mol. % substitution in $ZrO_2$, including the 3 mol. % $Y_2O_3$ in the starting material. The final samples for further characterization were produced by dipping each sample four times in the respective nitrate solution for 10 minutes per dip. In between immersions, the sample was lifted out from the solution, the surface of the pellet was wiped off with a piece of paper and the pellet was dried in a furnace at 200 °C in air for 10 minutes. The samples were thereafter sintered with a 10 °C/min ramp rate to 1500 °C followed by a 2-hour dwell at 1500 °C and 10 °C/min cooling (Zirkonofen 700 Vakuum, Zirkonzahn GmbH, Gais, Austria). The samples were weighed before the first dip (after pre-sintering) and after sintering, to calculate the quantity of infiltrated oxide, including a subtraction of the mass changes of a non-dipped 3YSZ reference sample before/after sintering. Additional samples with 1, 2, and 5 dips were also produced. A description of all the infiltration experiments is provided in section S1.

## 3.2. Optical measurements of translucence

Optical measurements were performed using a spectrophotometer (CM36-D, Konica Minolta, Japan) in reflection geometry. Measurements were performed using a white and a black background, a D65 light source and 8° angle of observation (n = 3). The translucency was evaluated by the translucency parameter (TP) provided as $TP = \sqrt{(L_b^* - L_w^*)^2 + (a_b^* - a_w^*)^2 + (b_b^* - b_w^*)^2}$, where subscript $b$ and $w$ denote black and white background, respectively. The lightness $L*$ and colour coordinates $a*$ and $b*$ in the $L*a*b*$ colour system was calculated by the Colibri software (Datacolor) using the CIE1976 10° supplementary standard observer.

## 3.3. Mounting, polishing, Knoop hardness and scanning electron microscopy

All samples where mounted in two-component acrylic, ClaroCit, in teflon molds (Ø 25 mm). Grinding and polishing where performed a semi-automatic Planpol using a 6 sample holder (Struers ApS, Ballerup, Denmark) with the following equipment in the given order: #220 MD-Piano with water (until plane), #1200 MD-Piano with water (~5 min), MD-Largo with 9 µm diamond suspension (~10 min), MD-Dac with 3 µm diamond suspension (~6 min), MD-Nap with 1 µm diamond suspension (~1 min), MD-Chem with 0.25 µm OP-S silica suspension (~1 min). Knoop hardness was measured with 1 kg of pressure (HK1) with a dwell time of 10 s using a Duramin 40 (Struers ApS, Ballerup, Denmark). Scanning electron microscopy (SEM) and energy dispersive X-ray spectroscopy (EDX) was performed using a Hitachi TM4000plus with an accelerating voltage of 10 kV and 5'000 times magnification for SEM images and 15 kV and 400 times magnification for EDX quantification.

## 3.4. Synchrotron radiation X-ray diffraction

### 3.4.1. Data collection

The sample were measured with synchrotron radiation X-ray diffraction at BM31 at SNBL, ESRF, France. The samples were discs (~0.5 mm) measured in transmission geometry, with the thickness compromising the beam absorption and obtaining reliable data from a volume of sample. This allowed the sample to be intact as after annealing, thus avoiding the formation of monoclinic YSZ phases that typically form during mortaring. The samples were mounted in a perforated plate, so that only the sample disc was present in beam.

The datasets were collected with three different detector distances, 250 mm (up to Q = 12.75 $Å^{-1}$), 700 mm and 970 mm , using a PILATUS3 X CdTe 2M detector. A Si single crystal air-bearing liquid nitrogen double-crystal monochromator was used, and the X-ray wavelength was 0.25509 Å. Due to the type of monochromator, a low intensity of $\lambda_3 = \lambda/3 = 0.08503$ Å was present in a small concentration in the beam. Due to the sample thickness and corresponding absorption and of the main wavelength beam, the data set was "polluted" by the diffraction signal from the $\lambda/3$ beam. Several additional peaks were present at low angles, which were removed by a Rietveld refinement fit using a python script and the GSAS-II software (section S3.1).

### 3.4.2. Rietveld refinements

The final Rietveld refinements were performed using TOPAS and inp files generated in Visual Studio code using the Topas editor plugin from Durham University [53,54]. All three detector distances were simultaneously included in the refinement for each material. The instrumental peak profile for each detector distance was determined using Si and the TCHZ peak function along with a simple axial model for peak asymmetry (mainly influencing the data from the 250 mm detector distance). Grain size and strain were used for additional peak broadening. A correction for sample displacement along the beam direction (debye-scherrer_specimen_displacement.txt from Topas Wiki [53]) and a background polynomial were included in all the refinements.

All the Rietveld refinements originated from the acknowledged tetragonal (t, t', t'') and cubic (c) structures of Zirconia, in line with previous reports from Krogstad et al. [47]. Although the phases (t, t', t'' and c, Figure 1) contain different quantities of Y (Table S6), the Y and Zr X-ray scattering cross sections are approximately equal. Although the scattering cross sections for Mg and Sc are different from that of Y and Zr, we were unable to extract any info regarding the content of Mg and Sc in the different phases. The analysis thus focuses only on the phases themselves.

To obtain a reliable starting point, the data sets were first fitted using the *t* and *t''* (*t'* for 0h samples) phases with fixed unit cell parameters, before adjusting with the *c* phase if needed. The fit was not improved by including all the four *t*, *t'*, *t''* and *c* phases. By including the *c*-phase early in the refinement, it could be overestimated and block a good fit for *t''*. After the initial fit refining the phase fractions, the unit cell parameters, strain, grain size and thermal displacement parameters were fitted as well. The *t*, *t'*, *t''* and *c*-phases for all samples have different unit cell parameters and tetragonality and the samples must be compared accordingly, such as through the variables defined below.

# 4. Results and discussion

## 4.1. Sample preparation and immersion experiments

Pre-sintered discs of 3Y-YSZ were infiltrated in a total of seven different solutions containing Mg, Sc, Y and mixtures of the three (Table 1, section S1). We record the mass changes (Table 1) and assume that all mass increase is caused by infiltration of the oxide of the new element, corrected for a mass loss through the sintering process seen from a reference sample. The Mg- and Y-samples obtain an infiltration of 4.6-4.7 mol. % of dopant, near the attempted total of 10 mol. % (Table 1, left). The other samples obtain 1.8-3.1 mol. % of dopant by the infiltration. EDX analysis (Table 1, right) show slightly higher dopant quantities, a total of 12-13 mol. % for Mg and Y, and around 9.5-10.5 mol. % for the rest.

The optimal infiltration protocol included wiping the surface of the pellet after dipping. Otherwise, surface scaling of dopant occurred, causing pieces of the surface to fall off and to re-dissolve during the next infiltration. Based on the EDX analysis we also believe that Y does not dissolve in significant quantities from the original 3YSZ material during infiltration of the other dopants.

*Table 1: Overview of mass changes and calculations of the quantity of infiltrated oxide in the samples from the final round of dipping. The samples with 4 dips are presented with further structural analysis below. The mass change is normalized for the mass loss.*

| Sample name | Mass change [wt. %] | Infiltrated oxide [mol. %] | Total mol. % not Zr | Mg [at. %] | Sc [at. %] | Y [at. %] | Mg+Sc+Y [at. %] |
|---|---|---|---|---|---|---|---|
| Mg-inf | 1.6 % | 4.6 % | 9.9 % | 6.7 % | - | 6.3 % | 13.0 % |
| Y-3YSZ | 4.5 % | 4.7 % | 10.3 % | - | - | 12.1 % | 12.1 % |
| Sc-3YSZ | 1.2 % | 2.0 % | 7.5 % | - | 2.7 % | 6.8 % | 9.5 % |
| MgY-3YSZ | 1.6 % | 2.4 % | 7.9 % | 2.1 % | - | 8.4 % | 10.5 % |
| MgSc-3YSZ | 0.8 % | 1.8 % | 7.3 % | 1.4 % | 1.7 % | 6.8 % | 9.9 % |
| ScY-3YSZ | 2.2 % | 2.9 % | 8.3 % | - | 1.5 % | 8.6 % | 10.1 % |
| MgScY-3YSZ | 1.9 % | 3.1 % | 8.5 % | 0.6 % | 1.3 % | 7.9 % | 9.8 % |

## 4.2. Structural characterization

### 4.2.1. 3YSZ and 5YSZ sintered for 0 and 2 hours

Two samples of 3YSZ and 5YSZ were prepared by sintering two samples at 1500 °C with 2 hours dwell with 10 °C/min and two samples with 0 min dwell and natural cooling. The samples with 0 min dwell and natural cooling represents the sample quenched from the state of reaching the sintering temperature.

Rietveld refinements show that the 3YSZ samples are dominated by the *t*-phase, with a small fraction of *t′* (Figure 2b-c). From 0h to 2h, a phase segregation occurs where the *t*-phase becomes more tetragonal and *t′*-phase reduces its tetragonality from 1.0095 to 1.0053, implying enrichment in Y and transformation of *t′* to *t″*, as indicated by the arrows in Figure 2a.

For 5YSZ, the sample is dominated by *t″* and the sample thus have a lower tetragonality than 3YSZ. 5YSZ goes through a phase segregation similar to 3YSZ, seen as increased tetragonality for *t* and reduced tetragonality for *t″* for 2h dwell compared to 0h dwell. For 5YSZ-0h, two refinement strategies can be used; using three tetragonal phases (*t*, *t′*, *t″*) of slightly different tetragonality (Figure 2d) or two tetragonal phases (*t*, *t''*) and the *c*-phase (Figure 2e). For 5YSZ-2h, *t*, *t''* and *c* combined gives the best fit.

Both refinement models for 5YSZ-0h lead to the same picture of the 0 h to 2 h evolution. The sample undergoes phase segregation into Y-rich and Y-depleted regions. The tetragonality deviation parameter captures this progressive phase segregation most clearly. It rises from 0.0019 to 0.0032 in 3YSZ between 0 h and 2 h, and from about 0.0030 to 0.0041 in 5YSZ over the same interval. The total tetragonality and the total pseudo-cubic volume behave differently. Both remain nearly unchanged within each composition, with variations of at most 0.0002 in tetragonality and 0.15 Å³ in volume.

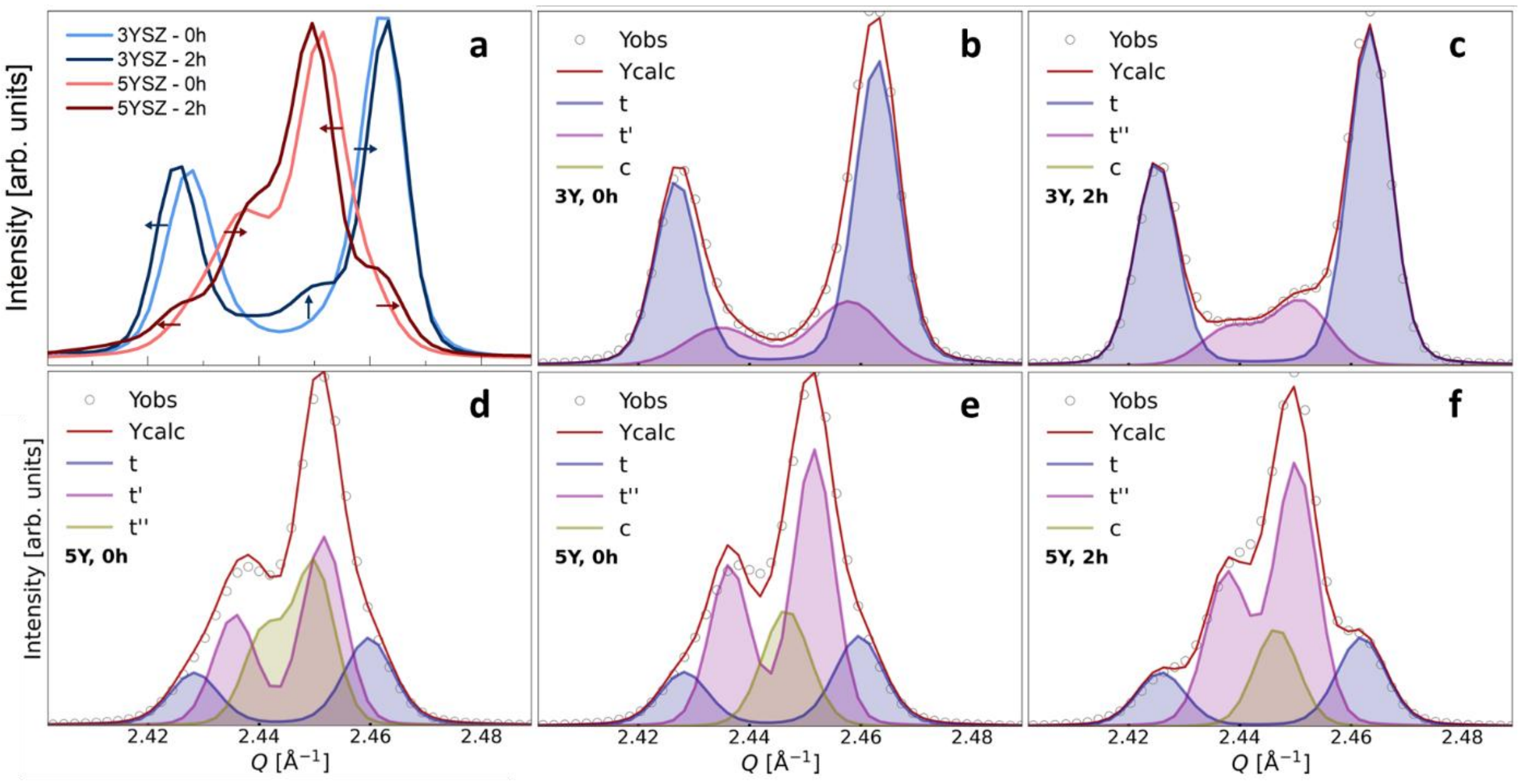

*Figure 2. Results from XRD and phase deconvolution from Rietveld refinement analysis of 3YSZ and 5YSZ in the region where the tetragonal (002) and (110) peaks and the cubic (200) peak appear. Raw data of a) 3YSZ-0h, 3YSZ-2h, 5YSZ-0h and 5YSZ-2h, and phase deconvolution of b) 3YSZ-0h, c) 3YSZ-2h, d) 5YSZ-0h analysed with three tetragonal phases, e) 5YSZ-0h analysed with two tetragonal phases and one cubic phase, and f) 5YSZ-2h. In a, the blue arrows show how the peaks change from 0 to 2 hours for 3YSZ, while the red arrows show how the peaks change from 0 to 2 hours for 5YSZ.*

*Table 2. Results from Rietveld refinements of the 3YSZ and 5YSZ samples annealed for 0 at 2 hours at 1500 °C. The table contains unit cell parameters, volume in the cubic unit cell, tetragonality* $\left(\frac{c_{tet}}{\sqrt{2}a_{tet}}\right)$ *and phase fractions for each phase, and the total tetragonality, tetragonality deviation and total volume. Figure 1 show the graphical difference in tetragonality between t, t' and t''. The volume (V) is given as the pseudo-cubic volume for the tetragonal samples* ($a^2 * c * 2$)*.*

| Sample names | 3YSZ – 0h | 3YSZ – 2h | 5YSZ – 0h (*t*, *t'*, *t''*) | 5YSZ – 0h (*t*, *t''*, *c*) | 5YSZ – 2h |
|---|---|---|---|---|---|
| **Rwp** | 4.85320 | 5.48725 | 6.60242 | 6.27613 | 7.22005 |
| **a [Å], *t*** | 3.6040(1) | 3.6034(1) | 3.6091(1) | 3.6076(1) | 3.6043(1) |
| **c [Å], *t*** | 5.1723(1) | 5.1765(1) | 5.1700(1) | 5.1681(2) | 5.1729(1) |
| **V [Å³], *t*** | 134.360(6) | 134.425(3) | 134.687(3) | 134.523(7) | 134.402(4) |
| **Tetragonality *t*** | 1.0148(1) | 1.0158(1) | 1.0129(2) | 1.0130(1) | 1.0148(1) |
| **a [Å], *t'*** | 3.6114(1) | - | 3.6211(2) | - | - |
| **c [Å], *t'*** | 5.1557(3) | - | 5.1546(2) | - | - |
| **V [Å³], *t'*** | 134.484(8) | - | 135.176(9) | - | - |
| **Tetragonality *t′*** | 1.0095(1) | - | 1.0066(10) | - | - |
| **a [Å], *t″*** | - | 3.6215(1) | 3.6235(1) | 3.6196(1) | 3.6217(1) |
| **c [Å], *t″*** | - | 5.1485(2) | 5.1435(2) | 5.1507(2) | 5.1479(1) |
| **V [Å$^3$], *t″*** | - | 135.048(6) | 135.069(6) | 134.964(6) | 135.044(4) |
| **Tetragonality *t″*** | - | 1.0053(1) | 1.0037(1) | 1.0062(1) | 1.0051(1) |
| **a [Å], *c*** | - | - | - | 5.1293(2) | 5.1291(1) |
| **V [Å$^3$], *c*** | - | - | - | 134.946(2) | 134.936(1) |
| **Phase fraction *t*** | 76.2(2) | 81.4(1) | 24.4(2) | 24.6(3) | 24.2(3) |
| **Phase fraction *t′*** | 23.8(2) | - | 41.2(3) | - | - |
| **Phase fraction *t″*** | - | 18.6(1) | 34.4(3) | 54.4(4) | 55.3(4) |
| **Phase fraction *c*** | - | - | - | 21.(5) | 20.5(5) |
| **Total tetragonality** | 1.0136 | 1.0138 | 1.0071 | 1.0066 | 1.0064 |
| **Tetragonality deviation** | 0.0019 | 0.0032 | 0.0028 | 0.0031 | 0.0041 |
| **Total volume [Å$^3$]** | 134.3868 | 134.5382 | 135.0176 | 134.8499 | 134.8647 |
| **Volume deviation** | 0.0448 | 0.1892 | 0.1622 | 0.1622 | 0.2251 |

### 4.2.2. Structural characterization of liquid infiltration samples

3Y samples were infiltrated with the dopants Mg(II), Sc(III), Y(III), and all combinations of the three. The XRD pattern of the Y-infiltrated sample (Figure 3) is similar to that of the 5YSZ 2h sample (Figure 2), implying that the Y-infiltration has changed the sample from 3YSZ to approximately a 5YSZ sample, in line with the mass changes in Table 1. The XRD patterns of the Mg- and Sc-infiltrated samples are clearly different from each other and different from the pure Y-doped samples (Figure 3). Still, the samples infiltrated in multi-element solutions combing Mg, Sc and Y (Table 1) show a diffraction pattern visually similar to the diffraction pattern of its individual components (Figure S5). Implying that each element has its unique effect on the atomic structure of the sample.

From Rietveld refinements using the same strategy as for the 3YSZ and 5YSZ samples, all the samples were fitted well using the *t*, *t″* and *c* phases with some variation in unit cell parameters. The output is reported in Table 3 and the phase deconvolutions in Figure 3. The key observations are:

- All samples are multiphase with significant degree of phase partitioning
  - The tetragonality of the *t*-phase is between 1.0153 and 1.0168 for all samples
  - The tetragonality of the *t″*-phase is between 1.0051 and 1.0066 for all samples
- The Y-infiltrated sample contains more *c* compared to 5YSZ 2h (60 % vs 21 %) and has a correspondingly lower total tetragonality compared to 5YSZ 2h. This occurs despite a similar total Y-content in the samples.
- Sc-infiltration has a shift in the tetragonal (200) and (110) *t″*-peaks towards higher angles (Figure 3d), implying a low *t″*-volume compared to 5YSZ. In 5YSZ, the *t″* (200) and (110) peaks are symmetrically located between those from the *t*-phase.
- All the Sc-infiltrated samples have the highest fraction of t and the highest total tetragonality. Sc-infiltration has the highest total tetragonality, but still far below that of 3YSZ.

- Mg-infiltration is found to fit well with a single cubic phase with low volume compared to the other samples. The small peak at Q = 2.427 $Å^{-1}$ is likely from a *t*-like phase present in <1 %.
  - Notably, the full-width-half-maximum of the peaks in the Mg-sample is equally narrow as the Si sample used to determine the instrumental peak broadening.

*Table 3. Results from Rietveld refinements of all the infiltrated samples. Mathematical uncertainties are calculated from Topas and are shown in parentheses.*

| **Sample** | **Mg** | **Sc** | **Y** | **Mg/Sc** | **Mg/Y** | **Sc/Y** | **Mg/Sc/Y** |
|---|---|---|---|---|---|---|---|
| **Rwp** | 6.54123 | 5.91581 | 8.62032 | 7.06174 | 7.93647 | 6.26367 | 7.22005 |
| **a [Å], t** | - | 3.6090(1) | 3.6059(1) | 3.6062(1) | 3.6082(1) | 3.6070(1) | 3.6079(1) |
| **c [Å], t** | - | 5.1825(1) | 5.1773(2) | 5.1838(1) | 5.1884(2) | 5.1795(2) | 5.1843(2) |
| **Pseudo-cubic volume [Å³], t** | - | 135.005(2) | 134.637(1) | 134.825(2) | 135.097(1) | 134.772(1) | 134.965(1) |
| **Tetragonality, t** | - | 1.0154(1) | 1.0153(1) | 1.0165(1) | 1.0168(1) | 1.0154(1) | 1.0161(1) |
| **a [Å], t''** | - | 3.6185(1) | 3.6221(1) | 3.6158(1) | 3.6209(1) | 3.6209(1) | 3.6211(1) |
| **c [Å], t''** | - | 5.1513(2) | 5.1519(1) | 5.1445(2) | 5.1469(2) | 5.1510(2) | 5.1493(2) |
| **Pseudo-cubic volume [Å³], t''** | - | 134.900(1) | 135.181(0) | 134.516(1) | 134.958(1) | 135.070(1) | 135.041(1) |
| **Tetragonality, t''** | - | 1.0066(1) | 1.0058(1) | 1.0061(1) | 1.0051(1) | 1.0059(1) | 1.0055(1) |
| **a [Å], cubic** | 5.0983(1) | 5.1278(2) | 5.1339(1) | 5.1224(1) | 5.1289(1) | 5.1301(2) | 5.1294(2) |
| **Volume [Å³], cubic** | 132.518(1) | 134.835(2) | 135.313(1) | 134.407(1) | 134.917(1) | 135.012(1) | 134.956(2) |
| **Phase fraction, t** | - | 42.0(3) | 16.2(1) | 33.3(3) | 20.2(2) | 33.1(3) | 24.8(2) |
| **Phase fraction, t''** | - | 37.8(4) | 23.3(1) | 26.0(3) | 26.7(3) | 40.8(4) | 36.8(2) |
| **Phase fraction, c** | 99.62644 | 20.2(4) | 60.5(1) | 40.7(3) | 53.1(3) | 26.1(4) | 38.4(3) |
| **Total pseudo-cubic volume [Å³]** | 132.525 | 134.929 | 135.172 | 134.573 | 134.963 | 134.954 | 134.988 |
| **Total tetragonality** | 1 | 1.0090 | 1.0038 | 1.0071 | 1.0048 | 1.0075 | 1.0060 |
| **Tetragonality deviation** | - | 0.0054 | 0.0046 | 0.0063 | 0.0051 | 0.0052 | 0.0050 |
| **Volume deviation** | - | 0.0616 | 0.1741 | 0.1660 | 0.0529 | 0.1229 | 0.0372 |

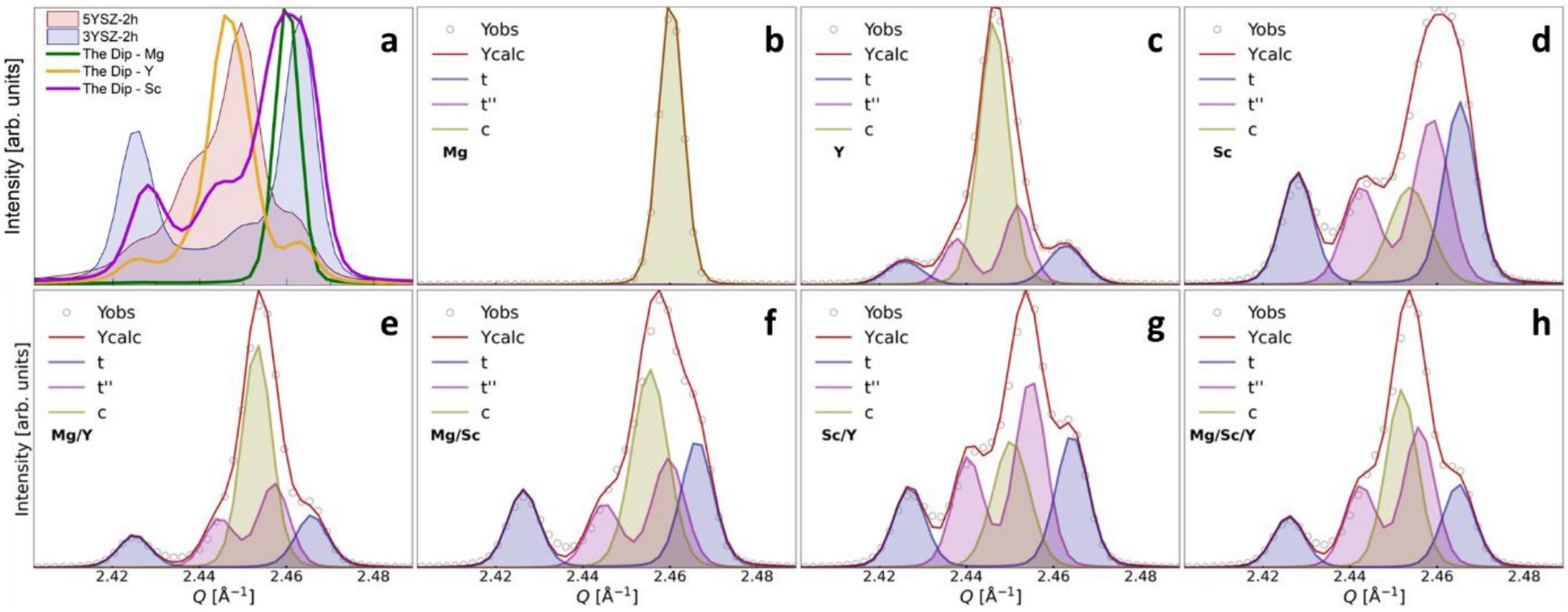

*Figure 3. a) XRD raw data comparing 3YSZ, 5YSZ, Y-infiltration, Sc-infiltration and Mg-infiltration. b-h) Phase deconvolution from the Rietveld refinement analysis of the infiltrated samples. The displayed figures are of the tetragonal (002) and (110)*

*reflection and the cubic (311) reflection from the longest detector distance (970 mm. The sample infiltration elements are provided in the figure.*

## 4.3. Grain size, hardness and translucency

### 4.3.1. Grain size

Grain size was evaluated qualitatively with SEM. Representative images are shown in Figure 4 for all samples. We have differentiated the samples into three groups: small, medium and large grain size, see Table 4. 3YSZ serves as benchmark for the small grain group, while 5YSZ belongs in the medium grain size group. The dark spots in the images are from pullouts from the polishing, but we still find the results representative with respect to the grain size analysis. Notably, the Y-infiltration sample show large grain size, but regions of small grain size can also be observed in the image, implying grain growth from infiltration coinciding at selected locations in the sample.

The most evident correlation to the grain size is the quantity of dopant (Table 4). The three samples with "large" grain size are three samples with the highest mol. % of dopant (Y + infiltration). Despite the equal starting point for all the samples before infiltration, it is interesting that the co-doping affects the final grain size in such different ways, exemplified by the observations below:

- The larger grain size of ScY- compared to MgScY-infiltration is caused only by the addition of Mg to the infiltration solution. However, pure Mg-infiltration alone has a large grain size.
- MgSc-infiltration has medium grain size while having the lowest infiltration quantity.
- MgScY and MgY have small grain size and simultaneously a large phase fraction of the *c*-phase

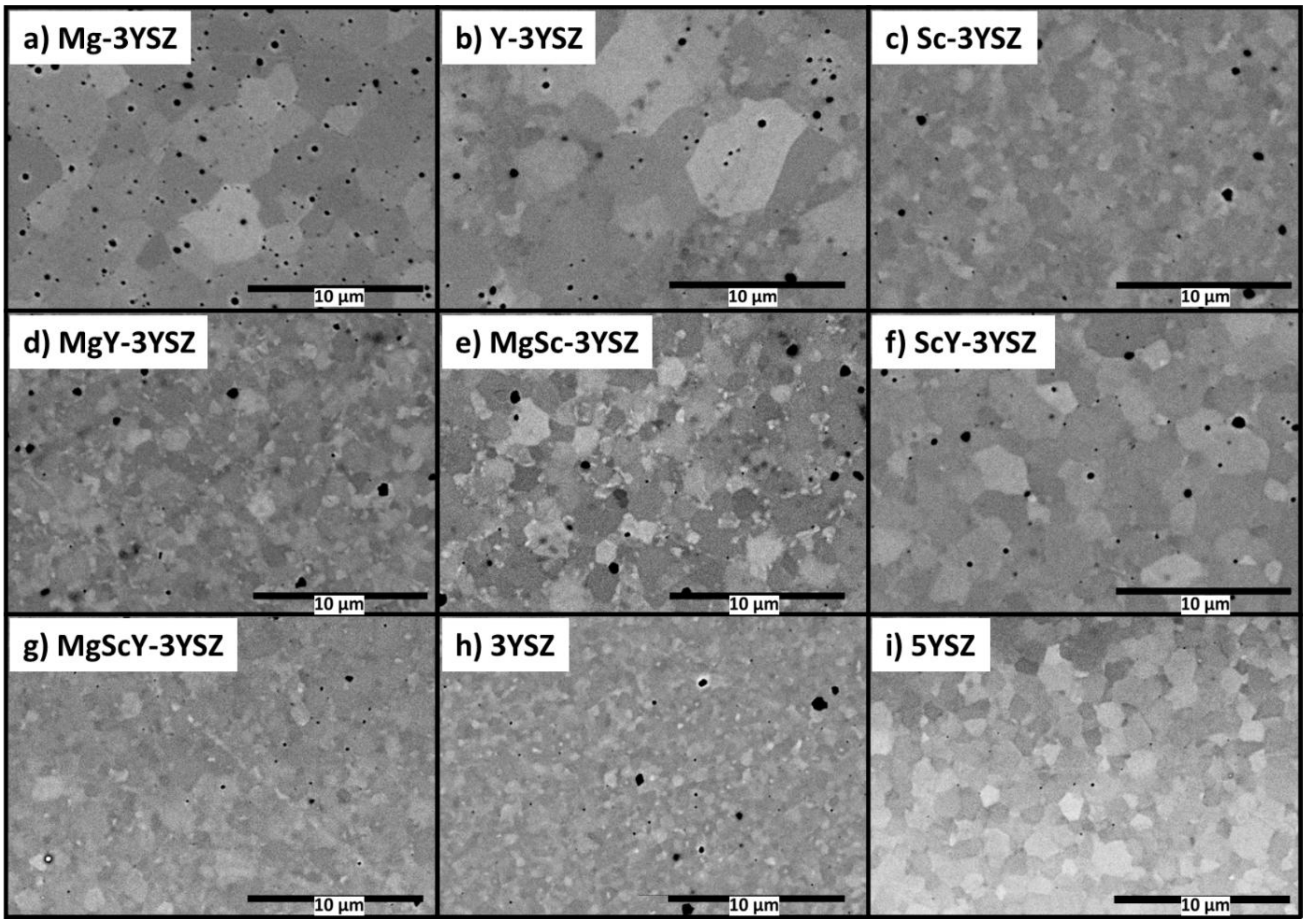

*Figure 4. SEM images of all samples acquired at 10 keV and at 5000x magnification. A black scale bar is set to 10 µm for all images and sample name is given in the figure.*

*Table 4. Table summarizing the findings across analyses. Grain size estimation (large/medium/small), translucency parameter, hardness, total dopant concentration, phase fraction of t and phase fraction of c. Samples sorted according to grain size.*

| Sample name | Grain size | Translucency parameter (TP) | Hardness (HK1) | Total mol. % not Zr | Phase fraction t | Phase fraction c |
|---|---|---|---|---|---|---|
| Y-infiltration | Large | 9.6 | 1280 | 10.0 | 17 | 60 |
| Mg-infiltration | Large | 9.2 | 1271 | 9.9 | 0 | 100 |
| 5YSZ | Medium | 15.8 | 1247 | 10.1 | 39 | 21 |
| ScY-infiltration | Medium | 13.0 | 1313 | 8.3 | 33 | 26 |
| MgSc-infiltration | Medium | 12.6 | 1255 | 7.3 | 34 | 37 |
| MgScY-infiltration | Small | 12.5 | 1335 | 8.5 | 25 | 40 |
| MgY-infiltration | Small | 12.5 | 1322 | 7.9 | 21 | 50 |
| Sc-infiltration | Small | 11.9 | 1314 | 7.5 | 42 | 21 |
| 3YSZ | Small | 13.1 | 1274 | 5.7 | 81 | 0 |

### 4.3.2. Knoop Hardness and translucency

Knoop hardness (HK1) with 1 kg pressure (Table 4, Figure 5) and translucency parameter (TP) measurements (Figure 5) were performed to place the results in a dental perspective, as these are the key material properties for dental materials. It is clear the Knoop Hardness test does not properly represent the mechanical properties of the samples. However, all the samples show a similar or higher HK1 compared to 3YSZ and 5YSZ. The Sc-, MgY-, ScY- and MgScY-infiltration samples show a higher hardness compared to the rest of the samples. Common for these four samples are moderate infiltration quantities (7.5-8.5 mol. % of not Zr) and a small or medium grain size (Figure 4).

TP was highest for the 5Y sample, and five samples (Sc-, MgY-, MgSc-, ScY- and MgScY-infiltration) displayed similar TP as 3Y. The two remaining samples (Mg-, Y-infiltration) showed lower TP than 3Y. The TP is low for samples with a high infiltration quantity and large grain size. There is also no correlation to the *c*-phase quantity. Enhanced optical properties was not achieved by the additional liquid infiltration doping, contrary to the well-known relationship between 3YSZ and 5YSZ [23]. More samples and focus on reproducibility of infiltrations can better resolve effect on optical properties. The changes seen in atomic- and grain structure imply that new understanding of substituted Zirconia is available by going beyond the classical single dopant approach and further explore co-doping.

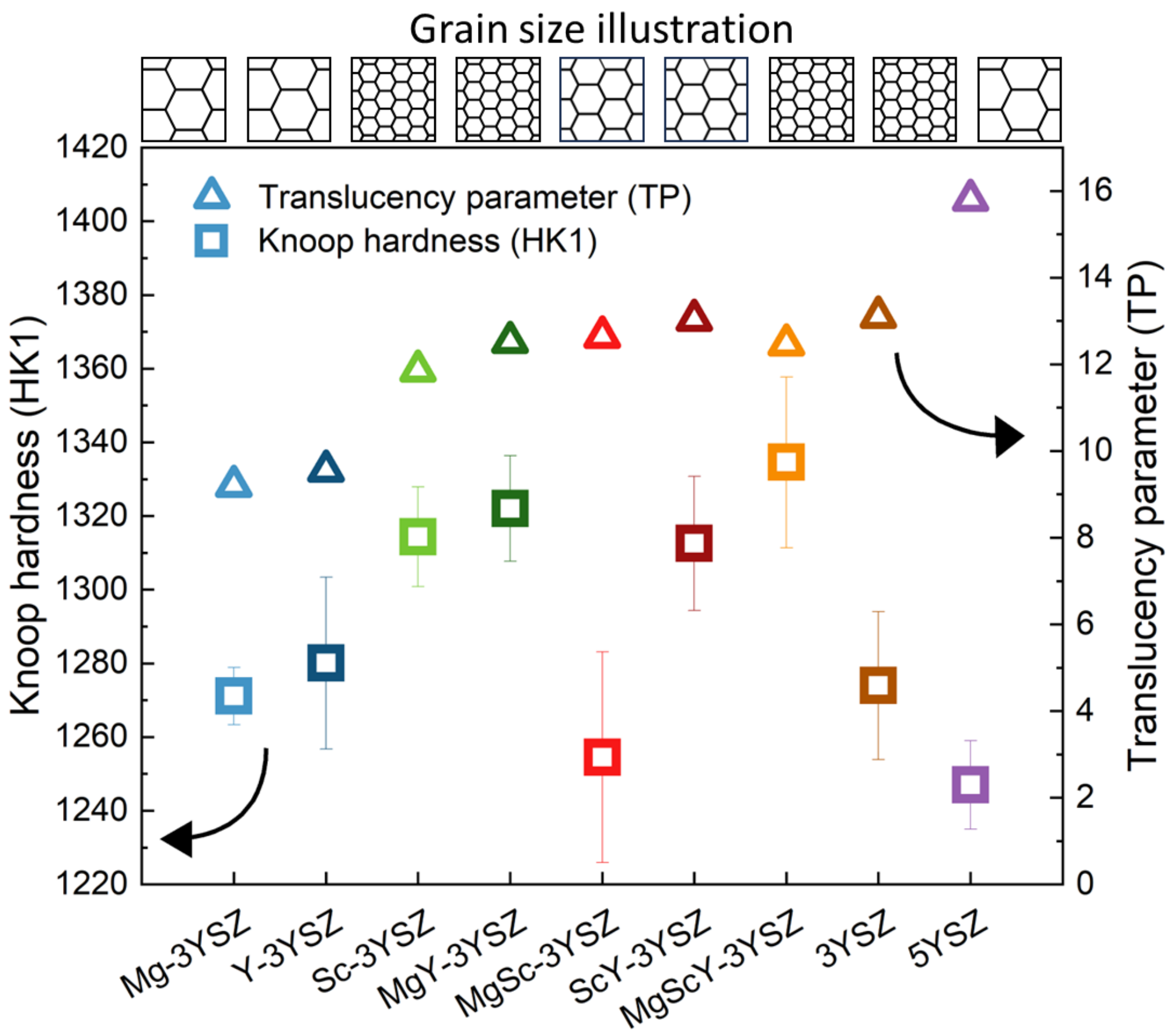

*Figure 5. Hardness Knoop 1 kg (HK1) on the left axis and translucency parameter (TP) on the right axis. Grain size illustration is provided for each sample on the top of the figure, indicating the small, medium or large grain size.*

# 5. Discussion

## 5.1. Structural aspects

Liquid infiltration has successfully altered the composition and atomic structure of the infiltrated 3Y-Zirconia, without forming binary oxide impurities ($Y_2O_3$, $Sc_2O_3$, MgO). This conclusion is derived from transmission XRD-data probing the sample volume, which shows that the dopants have reacted with the 3YSZ precursor throughout the sample. This is evident from the single cubic phase of the Mg-infiltration, which has a narrower XRD peak width than the Si-reference, implying a high crystallinity and homogeneity. The total doping level for Mg-infiltration is similar to 5Y, but the Mg-dopant introduces twice as many oxygen vacancies (Table S4) to become fully stabilized cubic Zirconia. Clearly, when the cubic phase is thermodynamically stable, sufficient interdiffusion for a homogeneous phase is achieved using the infiltration technique.

All samples beyond Mg-infiltration went through phase segregation into two-three phases during sintering, which is intensified by both sintering time and infiltration. In the case of 5YSZ-0h, we could fit the data with either *t*, *t′* and *t″*, or *t*, *t′′* and *c*, obtaining similar fit quality. Both models are reasonable; however, a largely different Y-content is calculated by use of the tetragonalities and Krogstad equations [55] (section S3.5) in the two cases, 10.7 vs 9.4 % (Table S7). Furthermore, the 3YSZ-0h sample was fitted with t and t′, while the 3YSZ-2h sample was fitted with t and t′′. The clear conclusion is that the Y-segregation is moving across the various phases, and the observed peaks are a result of inhomogeneous Y-concentration (or other dopants) at various locations, and that these are not actually different phases.

For the infiltrated samples, a much larger quantity of cubic phase is formed compared to the 3YSZ and 5YSZ samples, such as 60 % cubic phase for Y-infiltration compared to 21 % for 5YSZ-2h. This is quite surprising considering that the Y-content is similar in Y-infiltration (10.3 %) and 5YSZ (~10 %). Although EDX implies some excess Y in Y-infiltration compared to 5YSZ, calculating the Y-content from the equations by Krogstad *et al.* [32] provides a total of 9.9 at. % Y in Y-infiltration and 9.5 % Y in 5YSZ 2h (Table S7). Also, the *t*-phase has a higher tetragonality for all the infiltrated samples compared to 5YSZ (1.0153-1.0168 vs 1.0148), implying more phase segregation in the infiltrated samples. Since the infiltration applies dopant on the particle surface in the microporous green body, it accelerates the phase segregation and formation of Y-rich/dopant-rich phase at or near the grain boundaries [48]. In our case, excess dopant is present on the particle surfaces and in the grain boundaries already before sintering. This drives the phase segregation and cubic phase formation, in collaboration with the time dependent effect seen from the 3YSZ and 5YSZ samples with tetragonality deviations of 0.0019 vs 0.0032 for 3YSZ (0h vs 2h), 0.0028/0.0031 vs 0.0041 for 5YSZ (0h vs 2h) and 0.0046 for Y-infiltration.

Viewing all the samples in a holistic perspective, a large cation radius, high oxygen vacancy concentration (calculated based on nominal dopant concentration) and a high cubic phase fraction correlate with a low total tetragonality, see Figure 6a. The oxygen vacancies (red line in Figure 6a) trends well with the total tetragonality, a reasonable finding considering that the mechanisms for tetragonal distortion originates in displacements in the oxygen sublattice that subsequently affect the cation sublattice [48,56]. Several of the Sc-containing samples are located towards the right of Figure 6a, with a high tetragonality (1.0090), although not at the level of 3Y (1.0138).

Despite the strategy of small ionic radii, the increased dopant concentration has still led to a reduction in tetragonality. The small ionic radii dopant strategy has, however, resulted in a clear reduction in the volume of the *t''* and *c* phases relative to *t*. This is clear from phase deconvolution in Figure 3, showing a shift of the *t''* and *c* peaks towards higher angles. By plotting the sample volumes vs tetragonality (Figure 6b), we can see that all of the samples doped only with Y have increasing volume from $t \rightarrow t' \rightarrow t'' \approx c$ (Figure 6b, blue lines), while the Sc and Mg infiltrated samples mainly show a reduction in volume from $t \rightarrow t'' \approx c$ (Figure 6b, red lines). Notably, the volume trends with cation radii, while tetragonality trend with oxygen vacancies, two separate effect caused by the same dopant. Although the absolute volumes are unreliable due to a high correlation between sample displacement and the unit cell parameters, the relative trend between the *t*, *t'*, *t''* and *c* are reliable as they are calculated from the same datasets. These results indicate that the infiltrated elements tend to end up in the *t''* and *c* phases, not in the *t*-phase formed at the core of the grains. This is consistent with the description of dopant diffusion to the grain boundaries and nucleation of a dopant-rich phase [45,48], which is accelerated liquid infiltration and specific to its elements.

Extraction of quantitative information on elemental distribution from XRD and the data in Figure 6b has proven difficult as several assumptions are needed. We need reasonable ionic radii to predict volume expansions, which from Shannon as 8-coordinated cations become severely overestimated as effects from oxygen vacancies are not considered. Insight into local effects of each dopant would be preferred. Estimating reasonable ionic radii according to previous work [28,57], which infer a lower radius than Shannon 8-coordinated ions [58] as a result of oxygen vacancies, provides infiltration levels of approx. 2-5 % for most infiltration samples (S3.8). We further believe that the infiltrated elements end up solely in the *t''* and *c*-phases, not in *t*. To further explore co-doping in Zirconia, more insight is needed into the local environment, preferably around each dopant, *e.g.* by use of EXAFS such as the work by Li et al. [26–28], total scattering analysis [59] or neutron diffraction for larger elemental contrast. Considering the wide applications of Zirconia, and the real feasibility of developing such a model, we recommend this for future investigations.

This could also help resolve the debate regarding the tetragonal vs cubic phases in Zirconia, especially needed for dental materials. For nearly cubic Zirconia, the tetragonal distortion (originating in the anion sublattice) is insufficient to cause a tetragonal distortion of the cation sublattice and are indistinguishable from a cubic phase by XRD. In this case, XRD becomes insensitive to the true nature of the atomic structure. The reader must be aware of this ambiguity in XRD analysis, likely present in other publications on Zirconia as well. The 3YSZ samples are clearly free of cubic phase, but the 5YSZ samples are well described also without a cubic phase in the refinements. 5YSZ likely suffers from a common misconception that the cubic phase is dominating. From the present XRD analysis, max 21 % of 5YSZ contains cubic phase, likely less considering the SR-XRD and TEM investigations of 4Y-zirconia by Krogstad *et al.* [45,48]. For simplicity, the *t''* and *c*-phases can together be considered as non-transformable Zirconia phases (cannot undergo $t \rightarrow m$ transformation). A severe Y-segregation is needed to obtain a cubic phase, such as that we observe in the Y-infiltration sample, where the liquid infiltration is promoting the phase segregation.

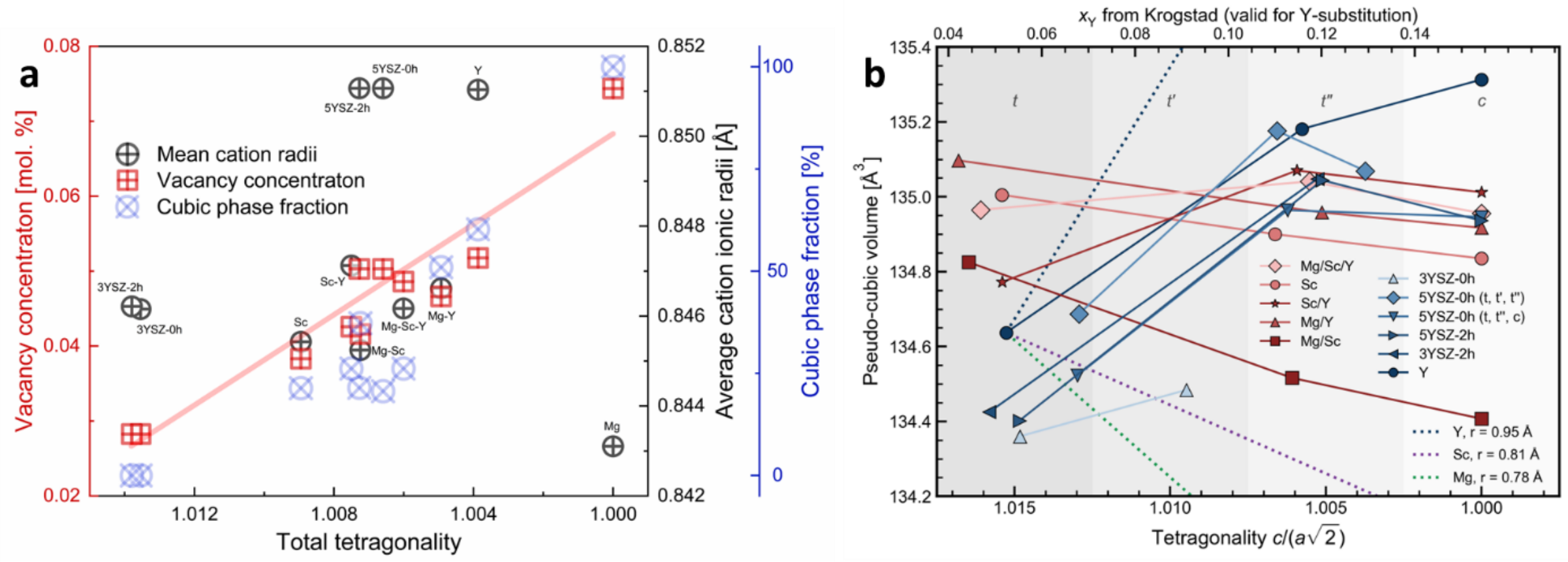

*Figure 6. a) Mean cation crystal radii [58] (black), oxygen vacancy concentration (red) and the cubic phase fraction (weak blue) plotted against total tetragonality. The samples names are written in the figure near the cation radii data point. The red line is the linear fit of the vacancy concentration. b) Pseudo-cubic volume of the Zirconia samples versus tetragonality. The top Y-axis show the Y-concentration as calculated from the Krogstad relation (S3.5). The dotted lines show the expected volume change from substitution of Y, Sc and Mg based on a simple model of ionic radii.*

## 5.2. Liquid infiltration of ceramics and dental prostheses

The applications of liquid infiltration to modify the atomic structure can reach applications within a broad range of the ceramic society. Various compositions can easily be made after the shaping step of the processing, which allow material testing in a realistic setting without spending time on the initial synthesis step for each composition. In dentistry, individual design of translucency and colouring can be performed after machining, making a realistic-looking restoration by use of translucency enhancing liquids [15]. Dental technicians are proficient in obtaining a realistic looking design and could benefit from additional tools when working with Zirconia. By selective design, translucency can be enhanced in certain regions, simultaneously as keeping the strength of the restoration largely unaffected by avoiding infiltration at the critical regions, such as the cervical margin and connectors [60]. The liquid can be applied on only the portion of the restoration that is less exposed to risk of fracture and where increased translucency will enhance aesthetics.

A key factor in the current results is the phase segregation and the large quantity of cubic phase observed in the infiltrated samples. Although 5YSZ and Y-infiltration have similar Y-content overall, the translucencies are not. This implies the need for mechanistic control of the liquid infiltration and sintering process. We believe liquid infiltration has the potential to make highly translucent Zirconia from a 3Y starting point, along with the authors of the published patent claims [17–20]. The key would be to control the phase segregation to make a more homogeneous material and less birefringent [23]. Besides that, grain size [23] and porosities [25,61] affect translucency, for which liquid infiltration can act as a sintering aid, helping in reducing pore volume.

Some developments are needed for achieving the desired level, both in engineering and scientific understanding. (i) Clinically relevant investigations to clearly state which clinical use cases and designs that can handle a reduction in strength [60], for the whole or portions of the restoration, without resulting in premature failure. (ii) The infiltration procedure, solvent and solution properties (*e.g.* rheology, optimized solvent evaporation, use of nanoparticles, etc.) must be optimized to allow for local enhancement of the translucency without unplanned effect on nearby areas. (iii) The interdiffusion process and phase segregation during sintering must be further explored, as the case of applying dopant in the grain boundaries before sintering is causing excessive phase partitioning.

# 6. Conclusion

This work show that liquid infiltration is a viable route to perform co-doping of Zirconia, and that it is applicable for several co-doping elements. It is a simple route to obtain compositions beyond the classical 3Y, 4Y, 5Y and other commercial materials. The co-doping successfully affects the atomic structure in a similar way as through conventional synthesis. Still, the fact that the dopant is by liquid infiltration applied by a form of coating on the 3YSZ particles may affect the final structure in a beneficial or undesired way, depending on the purpose. The tetragonality deviation of the liquid-infiltrated samples show that the phase segregation is slightly larger than that of 5YSZ sintered for 2 hours, and with excess cubic phase. The single phase of the Mg sample show that a homogeneous sample is achievable when favoured by thermodynamics. The liquid-infiltrated samples show a similar translucency compared to non-infiltrated 3YSZ, implying that the change in crystal structure alone was not sufficient for increasing the transmittance. A Knoop hardness at the level of 3YSZ or higher was observed for most samples. Grain growth occurs for all samples, to a higher or lower extent and without clear indications on why they behave differently. From these results we believe that new (or reinvented) Zirconia compositions can be identified that combine both a high strength, toughness and translucency. Furthermore, the liquid infiltration technique is likely applicable to several ceramics over a broad range of applications, for investigation of new materials or modifying current materials by such coating of large particles.

# 7. Declaration of generative AI and AI-assisted technologies in the manuscript preparation process

During the preparation of this work, the author(s) used the Anthropic AI-service Claude for help in searching, identifying and understanding patents to sufficiently review the previous work on translucency enhancing liquids, which is not fully covered in journal articles. The author(s) reviewed and edited the output as needed and take full responsibility for the content of the published article.

# 8. Acknowledgement

The authors would like to thank Freya Sofia Andersen (NIOM) for assistance in laboratory work, our colleagues at NIOM for scientific discussions, and the expertise of the staff at the Swiss-Norwegian beam lines (SNBL) at the European synchrotron radiation facility (ESRF). Funding: Ø.S.F. acknowledge funding from the Research Council of Norway (NFR) through projects 325345. The authors gratefully acknowledge SNBL, NFR project 208896.

# Supplementary

## S1. Infiltration experiments

The synthesis of YSZ samples for this study was performed by first following the commercial route for synthesis of 3YSZ until and including the pre-sintering step. Thereafter, to dope the 3YSZ with other elements, infiltration experiments by immersing the pellets in nitrate solutions were performed.

We have infiltrated 3YSZ pre-sintered pellets with Mg, Sc and Y and mixtures of the three. The first experiment consisted of performing 1-5 immersions (5 min) in an Mg, Y and 1:1 Mg:Y solution with intermediate drying for 5 minutes at 200 °C. The samples were weighed before immersion (after pre-sintering) and after sintering, see Table S1. We assume that all mass increase is caused by infiltration of an oxide of the new element, calibrated for a mass loss through the sintering process seen from a non-infiltrated sample (mass loss of 0.00126g).

A large weight mass are seen already for one immersion. Each system (Mg, Y, 1:1 Mg:Y) showed a mass increase with increasing amount of immersions, up to immersion no. 2-4, see Figure S1. However, a clear surface layer formed for several sample, especially for the Mg-samples. This surface layer causes flakes of the pellet to fall off after immersion no. 3 and the masses beyond this point cannot describe the Mg-content inside the pellet. A large portion of the mass increase is therefore likely on the pellet surface for several of these samples.

*Table S1. Overview of mass changes and calculations of the quantity of infiltrated oxide in the samples from the first round of infiltration experiments. The mass change is normalized for the mass loss observed in the samples that was not infiltrated (0.00126g).*

| # of immersions | Infiltration solution | Mass change [mg] | Mass change [wt. %] | Infiltrated oxide [wt. %] | Infiltrated oxide [mol. %] | Total mol. % not Zr | Mol. % of $Y_2O_3$ |
|---|---|---|---|---|---|---|---|
| **1** | Mg | 5.6 | 1.0 % | 1.0 % | 3.1 % | 8.5 % | 4.4 |
| **2** | Mg | 10.4 | 2.0 % | 2.0 % | 5.8 % | 11.1 % | 5.9 |
| **3** | Mg | 11.1 | 1.9 % | 1.9 % | 5.5 % | 10.8 % | 5.7 |
| **4** | Mg | 3.5 | 0.6 % | 0.6 % | 1.8 % | 7.3 % | 3.8 |
| **5** | Mg | -10.1 | -1.7 % | -1.8 % | -5.6 % | 0.4 % | 0.2 |
| **1** | 1:1 Mg:Y | 12.1 | 2.2 % | 2.1 % | 3.4 % | 8.8 % | 4.6 |
| **2** | 1:1 Mg:Y | 18.0 | 3.5 % | 3.4 % | 5.3 % | 10.6 % | 5.6 |
| **3** | 1:1 Mg:Y | 21.4 | 4.3 % | 4.1 % | 6.4 % | 11.6 % | 6.2 |
| **4** | 1:1 Mg:Y | 25.4 | 5.1 % | 4.8 % | 7.6 % | 12.7 % | 6.8 |
| **5** | 1:1 Mg:Y | 25.2 | 5.1 % | 4.9 % | 7.6 % | 12.7 % | 6.8 |
| **1** | Y | 19.8 | 3.9 % | 3.7 % | 4.0 % | 9.4 % | 4.9 |
| **2** | Y | 28.2 | 5.7 % | 5.4 % | 5.9 % | 11.1 % | 5.9 |
| **3** | Y | 37.9 | 6.8 % | 6.3 % | 6.8 % | 12.1 % | 6.4 |
| **4** | Y | 40.7 | 7.6 % | 7.0 % | 7.6 % | 12.8 % | 6.8 |
| **5** | Y | 34.0 | 6.7 % | 6.3 % | 6.8 % | 12.0 % | 6.4 |

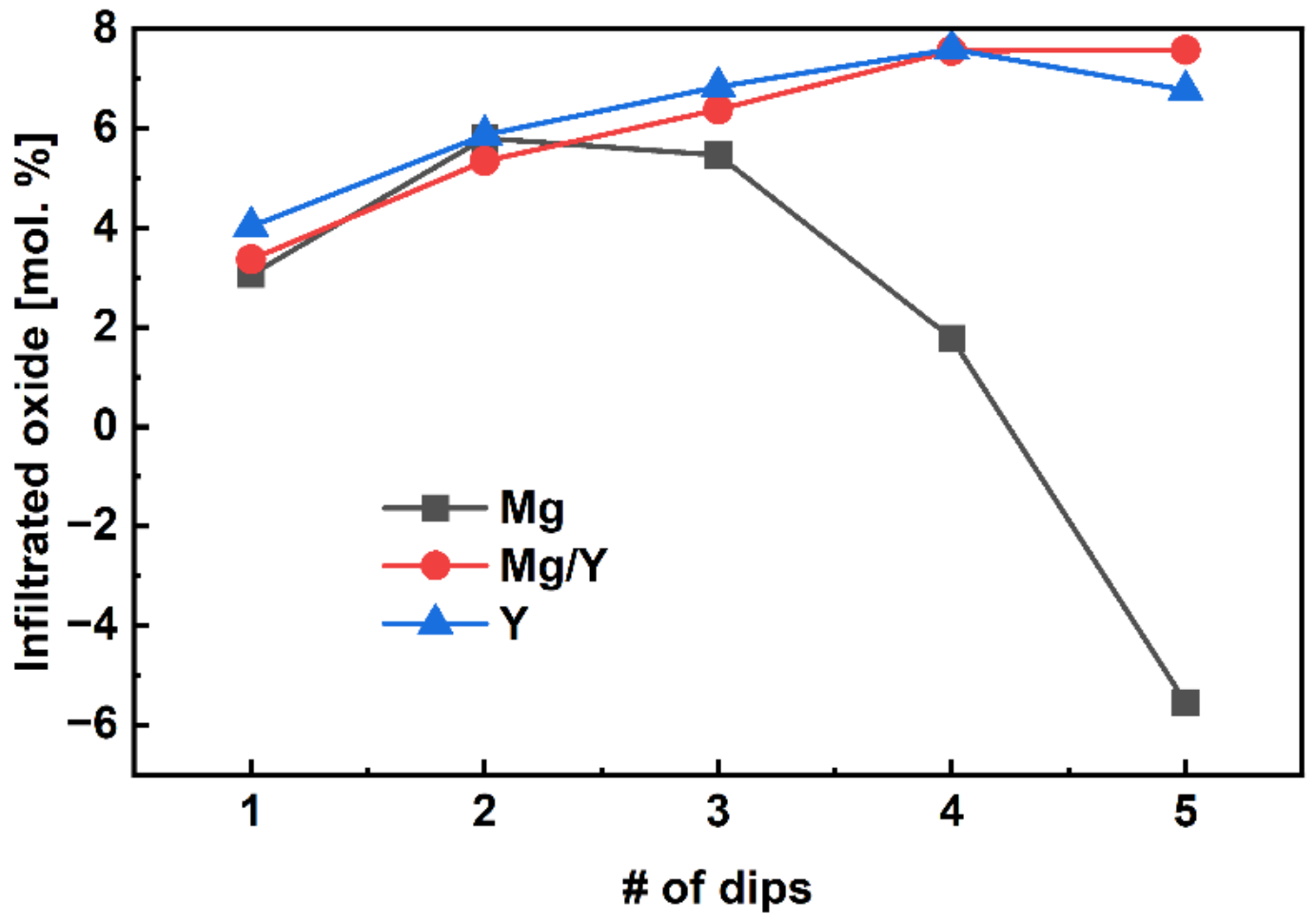

*Figure S1. Figure showing the evolution of the moles of infiltrated oxide as function of number of immersions. The Mg solution is showing a mass reduction due to flakes falling of the surface during the immersion procedure. The Mg/Y and Y solution show a mass increase up to immersion no. 4.*

In the second round of infiltration experiments, the surface of the pellet was dried with a piece of paper to ensure a high infiltration into the pellet core, and not only surface coverage. The pellets were immersed for 10 minutes, dried with paper and left to dry for 1 hour in air before sintering. This experiment included also single and mixed solutions with Sc (Table S2). The mass changes are lower than the experiment without drying the pellet surface, and a range of 0.2 – 2.6 mol. % of infiltrated oxide is obtained, see Table S2. The lowest quantities are infiltrated from the Mg, Sc and Mg-Sc mixed solutions.

*Table S2. Overview of mass changes and calculations of the quantity of infiltrated oxide in the samples from the second round of infiltration experiments, where the surface of the pellets were dried to remove nitrate solution from drying on the pellet surface. The mass change is normalized for the mass loss observed in the samples that was not infiltrated (0.00126g).*

| # of immersions | Infiltration solution | Mass change [mg] | Mass change [wt. %] | Infiltrated oxide [wt. %] | Infiltrated oxide [mol. %] | Total mol. % not Zr | Mol. % of $Y_2O_3$ |
|---|---|---|---|---|---|---|---|
| **2** | Mg | 0.2 | 0.1 % | 0.1 % | 0.2 % | 5.8 % | 3.0 |
| **2** | Y | 12.0 | 2.6 % | 2.6 % | 2.8 % | 8.2 % | 4.3 |
| **2** | Sc | 1.6 | 0.4 % | 0.4 % | 0.7 % | 6.3 % | 3.2 |
| **2** | 1:1 Mg:Y | 5.7 | 1.2 % | 1.2 % | 1.9 % | 7.4 % | 3.8 |
| **2** | 1:1 Mg:Sc | 1.6 | 0.3 % | 0.3 % | 0.7 % | 6.3 % | 3.3 |
| **2** | 1:1 Y:Sc | 6.7 | 1.4 % | 1.4 % | 1.9 % | 7.4 % | 3.8 |
| **2** | 1:1:1 Mg:Y:Sc | 4.8 | 1.1 % | 1.1 % | 1.8 % | 7.3 % | 3.8 |
| **1** | Mg | 1.3 | 0.3 % | 0.3 % | 0.8 % | 6.3 % | 3.3 |
| **1** | Y | 12.3 | 2.5 % | 2.4 % | 2.6 % | 8.1 % | 4.2 |
| **1** | Sc | 1.9 | 0.4 % | 0.4 % | 0.7 % | 6.2 % | 3.2 |
| **1** | 1:1 Mg:Y | 5.3 | 1.1 % | 1.1 % | 1.8 % | 7.3 % | 3.8 |
| **1** | 1:1 Mg:Sc | 1.4 | 0.3 % | 0.3 % | 0.7 % | 6.2 % | 3.2 |
| **1** | 1:1 Y:Sc | 5.2 | 1.1 % | 1.1 % | 1.5 % | 7.0 % | 3.6 |
| **1** | 1:1:1 Mg:Y:Sc | 3.4 | 1.0 % | 1.0 % | 1.6 % | 7.1 % | 3.7 |

In the third and final round of infiltration experiments the pellets were immersed 4-5 times each for 10 minutes. The pellet surface was dried with paper after immersion and before a 10 minutes heating at 200 °C. Finally the pellets were sintered at 1500 °C. The mass increases are higher and the quantity of infiltrated oxide range from 1.8 – 4.9 mol. %, see Table S3.

*Table S3. Overview of mass changes and calculations of the quantity of infiltrated oxide in the samples from the final round of infiltration experiments. The samples with 4 immersions are presented with further structural analysis below. The mass change is normalized for the mass loss observed in the samples that was not immersed (0.00126g).*

| # of immersions | Infiltration solution | Mass change [mg] | Mass change [wt. %] | Infiltrated oxide [wt. %] | Infiltrated oxide [mol. %] | Total mol. % not Zr | Mol. % of $Y_2O_3$ |
|---|---|---|---|---|---|---|---|
| **5** | Mg | 6.5 | 1.3 % | 1.3 % | 3.9 % | 9.3 % | 4.9 |
| **5** | Y | 23.2 | 4.8 % | 4.5 % | 4.9 % | 10.2 % | 5.4 |
| **5** | Sc | 5.0 | 1.1 % | 1.1 % | 2.0 % | 7.4 % | 3.9 |
| **5** | 1:1 Mg:Y | 8.1 | 1.7 % | 1.7 % | 2.7 % | 8.2 % | 4.3 |
| **5** | 1:1 Mg:Sc | 5.1 | 1.0 % | 1.0 % | 2.3 % | 7.8 % | 4.0 |
| **5** | 1:1 Y:Sc | 8.2 | 2.1 % | 2.0 % | 2.7 % | 8.1 % | 4.2 |
| **5** | 1:1:1 Mg:Y:Sc | 10.7 | 2.1 % | 2.1 % | 3.4 % | 8.8 % | 4.6 |
| **4** | Mg | 7.1 | 1.6 % | 1.6 % | 4.6 % | 9.9 % | 5.2 |
| **4** | Y | 21.4 | 4.5 % | 4.3 % | 4.7 % | 10.0 % | 5.3 |
| **4** | Sc | 5.5 | 1.2 % | 1.1 % | 2.0 % | 7.5 % | 3.9 |
| **4** | 1:1 Mg:Y | 7.0 | 1.6 % | 1.5 % | 2.4 % | 7.9 % | 4.1 |
| **4** | 1:1 Mg:Sc | 3.7 | 0.8 % | 0.8 % | 1.8 % | 7.3 % | 3.8 |
| **4** | 1:1 Y:Sc | 10.6 | 2.2 % | 2.1 % | 2.9 % | 8.3 % | 4.3 |
| **4** | 1:1:1 Mg:Y:Sc | 9.2 | 1.9 % | 1.9 % | 3.1 % | 8.5 % | 4.4 |

By comparing the two sets of samples immersed 1-2 and 4-5 times we see a clear trend in element absorption, see Figure S2. Figure showing the mass increase of the pellets infiltrated 1-4 times including drying the surface in air or with paper after the infiltration.. For the same element solution the mass gains are similar. There is an effect from more immersions, but in the set of 1-2 immersions the masses are similar, and in the set of 4-5 immersions the masses are similar. Between the two sets a main difference is also the drying. It seems like furnace drying at 200 °C allows a higher mass increase of infiltrated oxide.

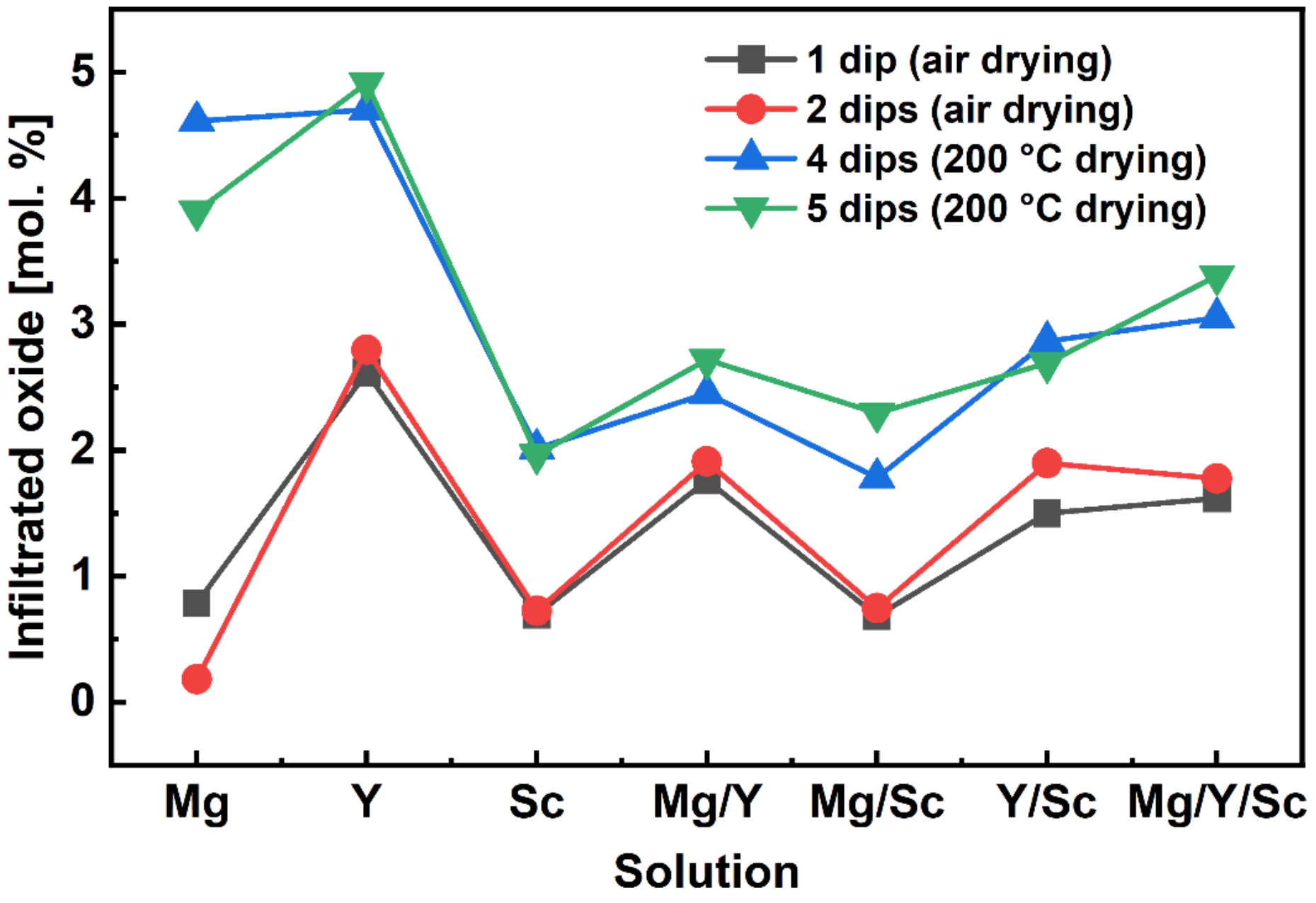

*Figure S2. Figure showing the mass increase of the pellets infiltrated 1-4 times including drying the surface in air or with paper after the infiltration.*

# S2. Expectations from substitutions and cation radii

*Table S4. Substitution equations (Kröger-Vink) and ionic radii for Zr(IV), Y(III), Sc(III) and Mg(II) [58] in 8-coordination.*

| Substitution equation (Kröger-Vink) | Ionic radii [Å], CN = 8 |
|---|---|
| ZrO2 | 0.84 |
| $Y_2O_3 \rightarrow 2Y_{Zr}^{\prime} + 3O_O^X + V_O^{\bullet\bullet}$ | 1.019 |
| $Sc_2O_3 \rightarrow 2Sc_{Zr}^{\prime} + 3O_O^X + V_O^{\bullet\bullet}$ | 0.87 |
| $\mathrm{MgO} \rightarrow Mg_{Zr}^{\prime\prime} + O_O^X + V_O^{\bullet\bullet}$ | 0.89 |

*Table S5. Overview with mol. % of dopants used to calculate the average ionic radii based on the Shannon radii in Table S4, the smaller ionic radii defined in section S3.8 and the average calculated oxygen vacancy concentration.*

| Sample name | mol. % Mg | mol. % Sc | mol. % Y | mol. % Y total | Mol. % infiltrated oxide | Mol. % dopant total | avg ionic radii, Shannon radii | avg ionic radii from section S3.8 | Avg vacancy concentration |
|---|---|---|---|---|---|---|---|---|---|
| 3YSZ_0h | | | | 0.0559 | | 0.0559 | 0.8500 | 0.8462 | 0.0280 |
| **3YSZ_2h** | | | | 0.0559 | | 0.0559 | 0.8500 | 0.8462 | 0.0280 |
| **5YSZ_0h** | | | | 0.1006 | | 0.1006 | 0.8580 | 0.8511 | 0.0503 |
| **5YSZ_2h** | | | | 0.1006 | | 0.1006 | 0.8580 | 0.8511 | 0.0503 |
| **Mg-3YSZ** | 0.0461 | | | 0.0534 | 0.0461 | 0.0995 | 0.8519 | 0.8431 | 0.0959 |
| **Y-3YSZ** | | | 0.0470 | 0.1003 | 0.0470 | 0.1003 | 0.8580 | 0.8510 | 0.0502 |
| **Sc-3YSZ** | | 0.0201 | | 0.0548 | 0.0201 | 0.0749 | 0.8504 | 0.8454 | 0.0475 |
| **MgY-3YSZ** | 0.0122 | | 0.0122 | 0.0668 | 0.0245 | 0.0791 | 0.8526 | 0.8466 | 0.0518 |
| **MgSc-3YSZ** | 0.0089 | 0.0089 | | 0.0549 | 0.0178 | 0.0728 | 0.8505 | 0.8452 | 0.0497 |
| **ScY-3YSZ** | | 0.0143 | 0.0143 | 0.0687 | 0.0287 | 0.0830 | 0.8527 | 0.8471 | 0.0487 |
| **MgScY-3YSZ** | 0.0102 | 0.0102 | 0.0102 | 0.0644 | 0.0305 | 0.0848 | 0.8523 | 0.8462 | 0.0576 |

# S3. Structural analysis

## S3.1 Removing the contribution from $\lambda_3$

Because the data sets contained a signal from $\lambda_3$, we wanted to remove the signal from the data to allow better analysis of the real data. The program used was GSAS-II in combination with Python (GSAS-II scriptable), and GSAS-II does not correctly fit two different wavelength when there is a large Table S 1, Figure S 1 difference in between the two wave length (i.e. when it does not originate from two different spectral lines). We there performed a Rietveld refinement with three tetragonal phases with $\lambda_3$ and subtracted this refinement from the original dataset. This provided an ok dataset with less contribution from the $\lambda_3$ peaks, and it was more obvious which peaks that originated from the $\lambda_3$ wavelength.

## S3.2 Definition of parameters used in structural analysis

To perform the structural analysis with a high accuracy, certain values are calculated from the data. The total tetragonality is calculated to provide the overall tetragonal character of the sample by weighting the tetragonalities of the different phase by the phase fractions, see equation below.

$$Total\ tetragonality = \Delta Tet = \sum_{all\ phases} Tetragonality_i * Phase\ Fraction_i$$

The total volume calculates the corresponding overall unit cell volume of the sample for the cubic unit cell settings. This value can better correlate samples with a large quantity of the cubic phase, but where the unit cell parameters of the cubic cell varies between the samples. The formula is provided below.

$$Total\ volume = \sum_{all\ phases} Volume_i * Phase\ Fraction_i$$

The tetragonality deviation is calculated to show how inhomogeneous the sample is. For a single-phase sample the tetragonality deviation is zero, while for a multi-phase sample the tetragonality deviation show to which extent the phases deviate from the total tetragonality. This allows for a direct comparison of the phase distribution between the samples.

$$Tetragonality\ deviation = \sum_{all\ phases} abs(\Delta Tet_i - \Delta Tet_{total}) * Phase\ Fraction_i$$

## S3.3 Cif-files

The Rietveld refinements were performed with already published cif-files as basis [62,63]. They contain different Y-content (see Table S6). However, due to the similar scattering factor of Y and Zr, the elements are indistinguishable in the refinements. For Sc and Mg, the scattering factors are different. However, the low Mg and Sc concentrations in the samples imply a very low effect on the intensities, likely below the resolution of our experiments. The refinements were therefore performed with a set quantity of Y as the substituent in the cif-files since we cannot know in which phase the Mg and Sc is. We believe this assumption has a negligible effect on the refinement outcome.

*Table S6. Concentration of Y-atoms in the three tetragonal phases and the cubic phase. Collected from reported cifs.*

| Phases | t | t' | t'' | cubic |
|---|---|---|---|---|
| **Y atoms at Zr site [%]** | 4.9 | 9.4 | 11.6 | 21.4 |

## S3.4 Visual analysis of the X-ray diffraction patterns

X-ray diffraction analysis was performed in detail for the samples dipped 4 times, that are listed in Table S3. All starting samples were 3YSZ and have been dipped in order to be infiltrated with more substituent elements. The data reveal a clear difference in structure depending on the dipping solution. We first compare the Y-infiltrated sample with 3YSZ and 5YSZ samples from Tosoh, see Figure S3. The Y-infiltrated sample contains a total of 11.1 mol. % $Y_2O_3$, very close to the 10.8 mol. % Y in 5YSZ from Tosoh. The XRD pattern of this sample is very similar to 5YSZ, indicating that the infiltration has successfully changed the sample from a 3YSZ to a 5YSZ sample.

The 3YSZ and 5YSZ samples have different intensities of the XRD reflections. The 3YSZ sample usually has a higher content of the t-phase, giving the two peaks at Q = 2.425 Å (002) and Q = 2.465 Å (110). For 5YSZ, the majority phase is t' (or t'' depending on the author), which has a lower degree of tetragonal distortion and the two peaks are spaced closer together. Both the 3YSZ and 5YSZ samples do in fact contain portions of each phase.

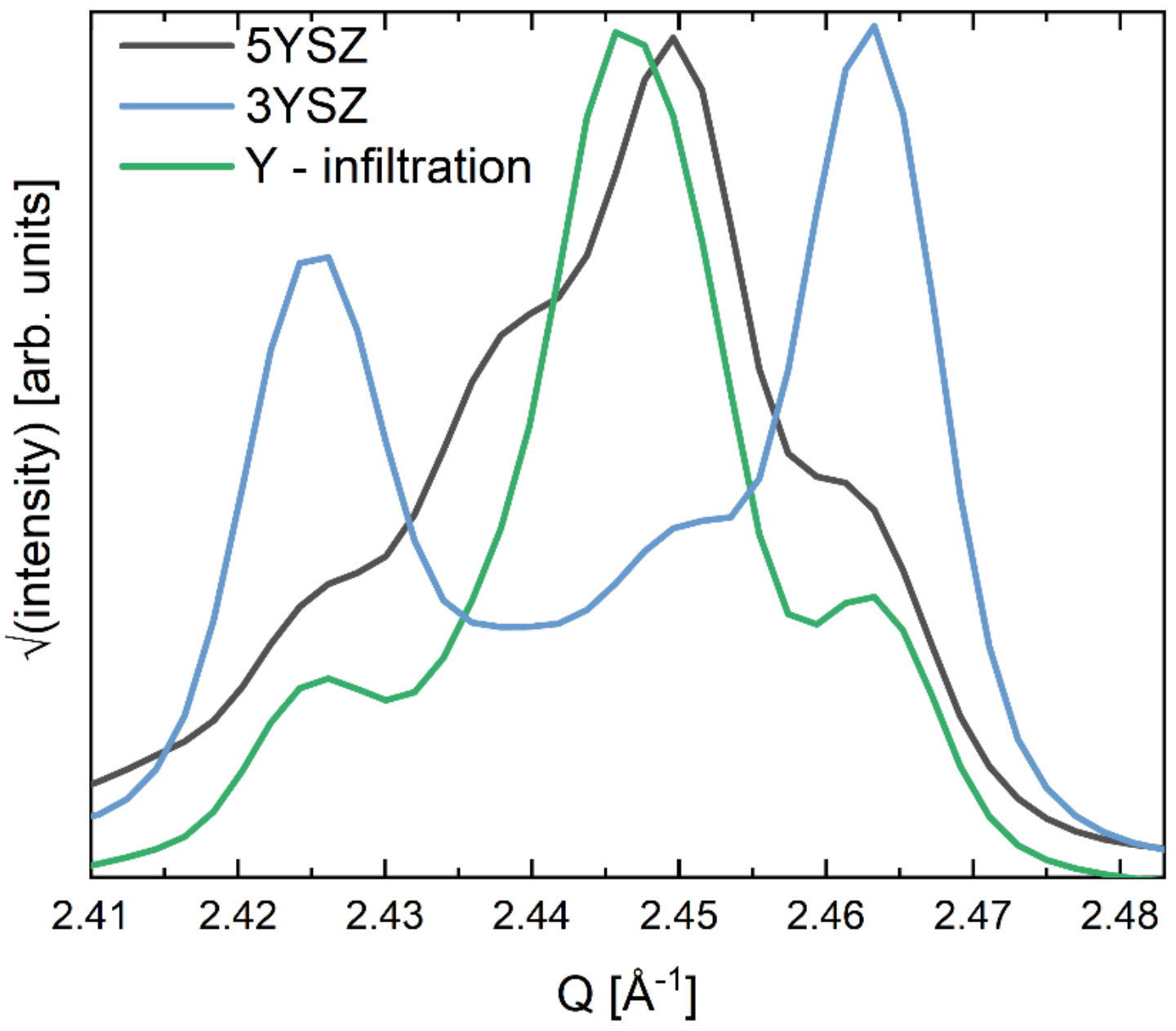

*Figure S3. Figure showing the second peak in the diffraction pattern for the YSZ-samples. The 3YSZ and 5YSZ samples are as produced from Tosoh, while the Y-infiltration sample is the 3YSZ samples that has been dipped to be infiltrated with more Y.*

The XRD pattern of the Mg-infiltrated and Sc-infiltrated samples have a clearly different look compared to the 3YSZ and 5YSZ samples, see Figure S4. The Mg-sample has a very strong peak at Q = 2.46 Å, near the location of the (110) reflection for the tetragonal t-phase, and only a very weak signal near the location of the (002) reflection of the t-phase. The Sc-sample also has a strong peak at Q = 2.46 Å, but it is much broader. In addition, the Sc-sample have peak intensity similar to the Y-infiltration sample, although with different intensities.

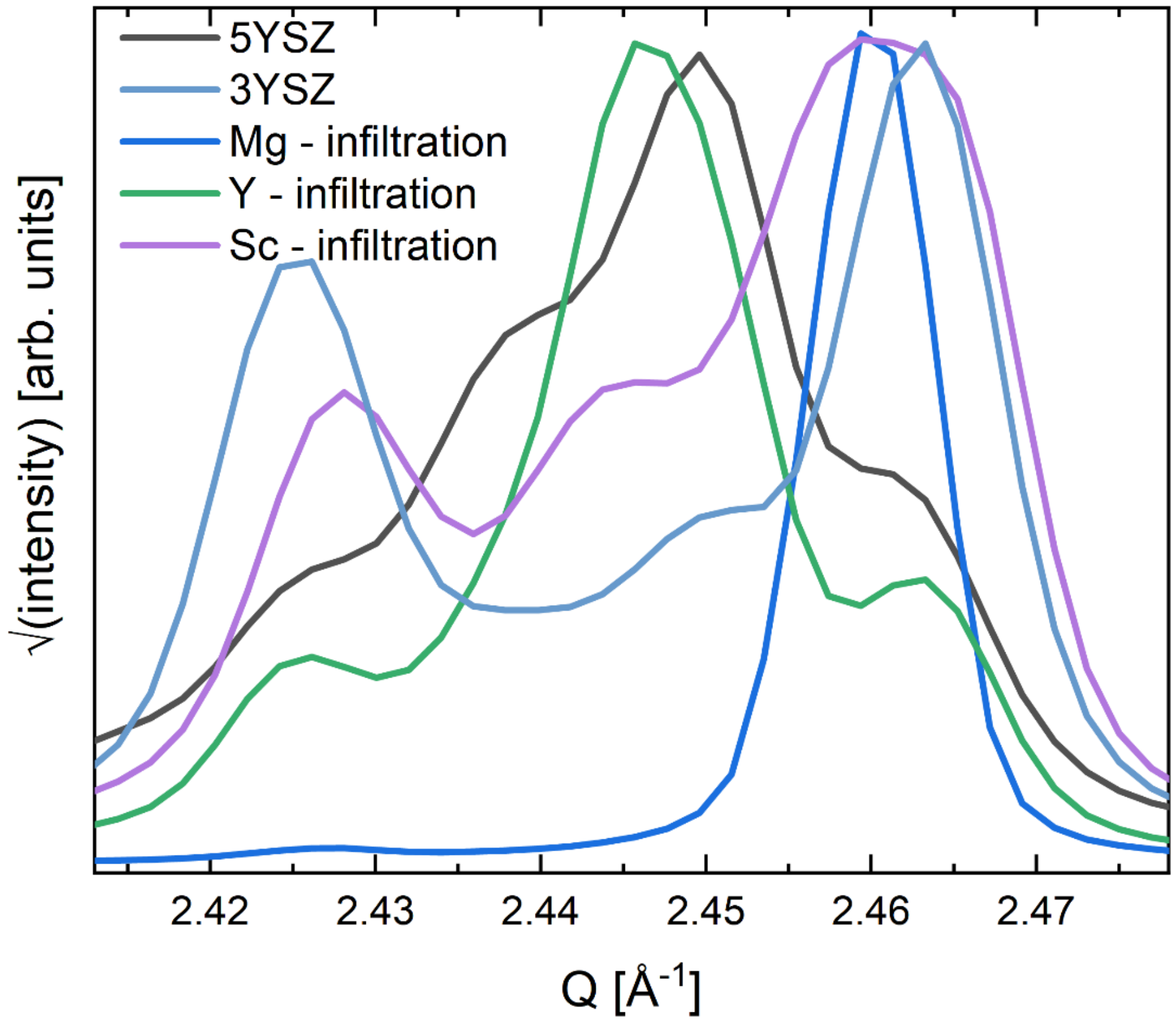

*Figure S4. Figure showing the second peak in the diffraction pattern for the YSZ-samples. The 5YSZ and 3YSZ samples are as produced from Tosoh, while the Mg-, Sc- and Y-infiltrated samples are produced by the dipping procedure described above.*

The three infiltrated samples are clearly distinct from each other, and the samples that have been dipped in multi-element solutions show a diffraction pattern similar to the diffraction pattern or its individual components, see Figure S5.

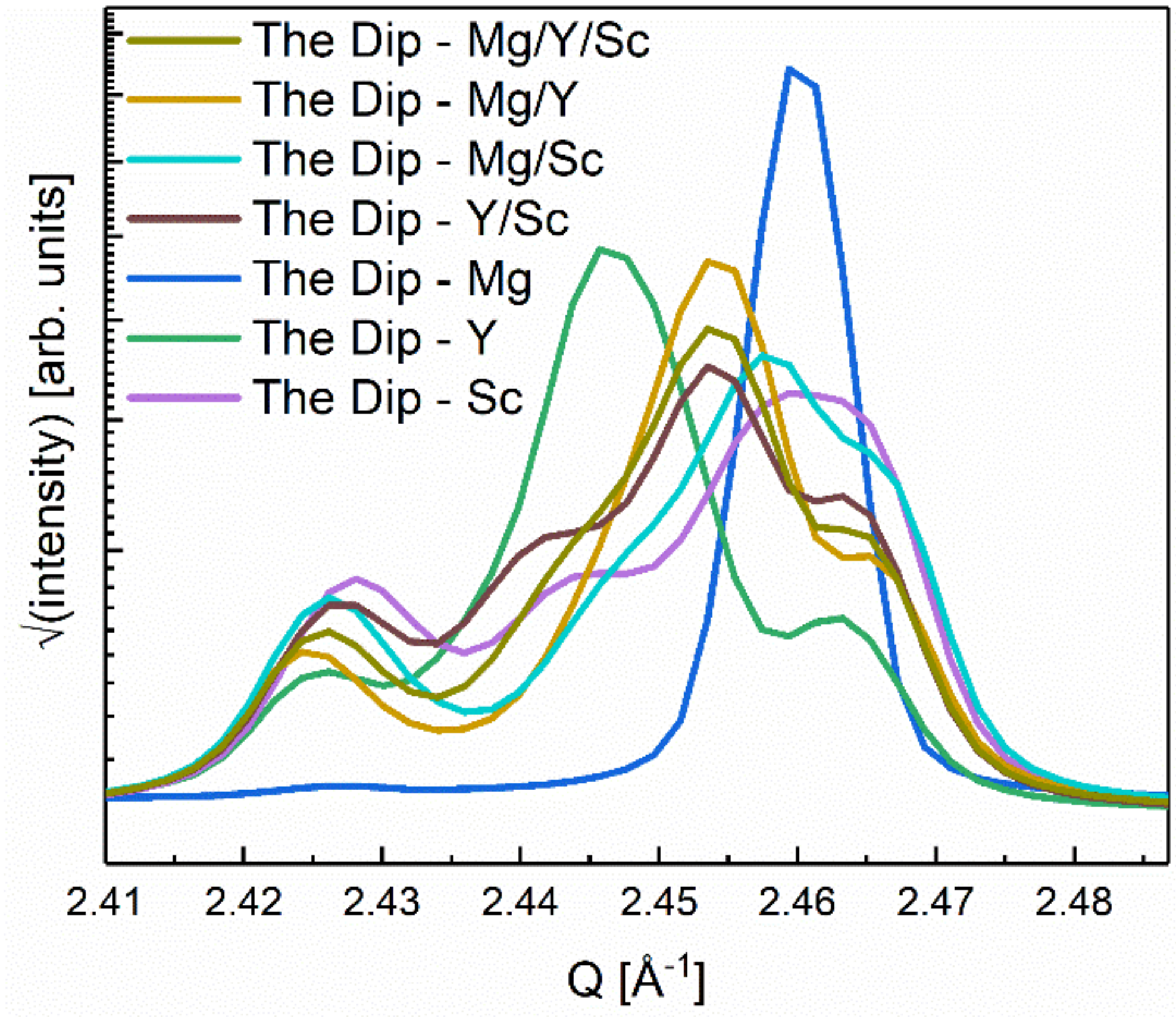

*Figure S5. Figure showing the second peak in the diffraction pattern for the YSZ-samples. The 5YSZ and 3YSZ samples are as produced from Tosoh, while the all the dipped samples are produced by the dipping procedure described above.*

## S3.5 Rietveld refinements of the XRD data

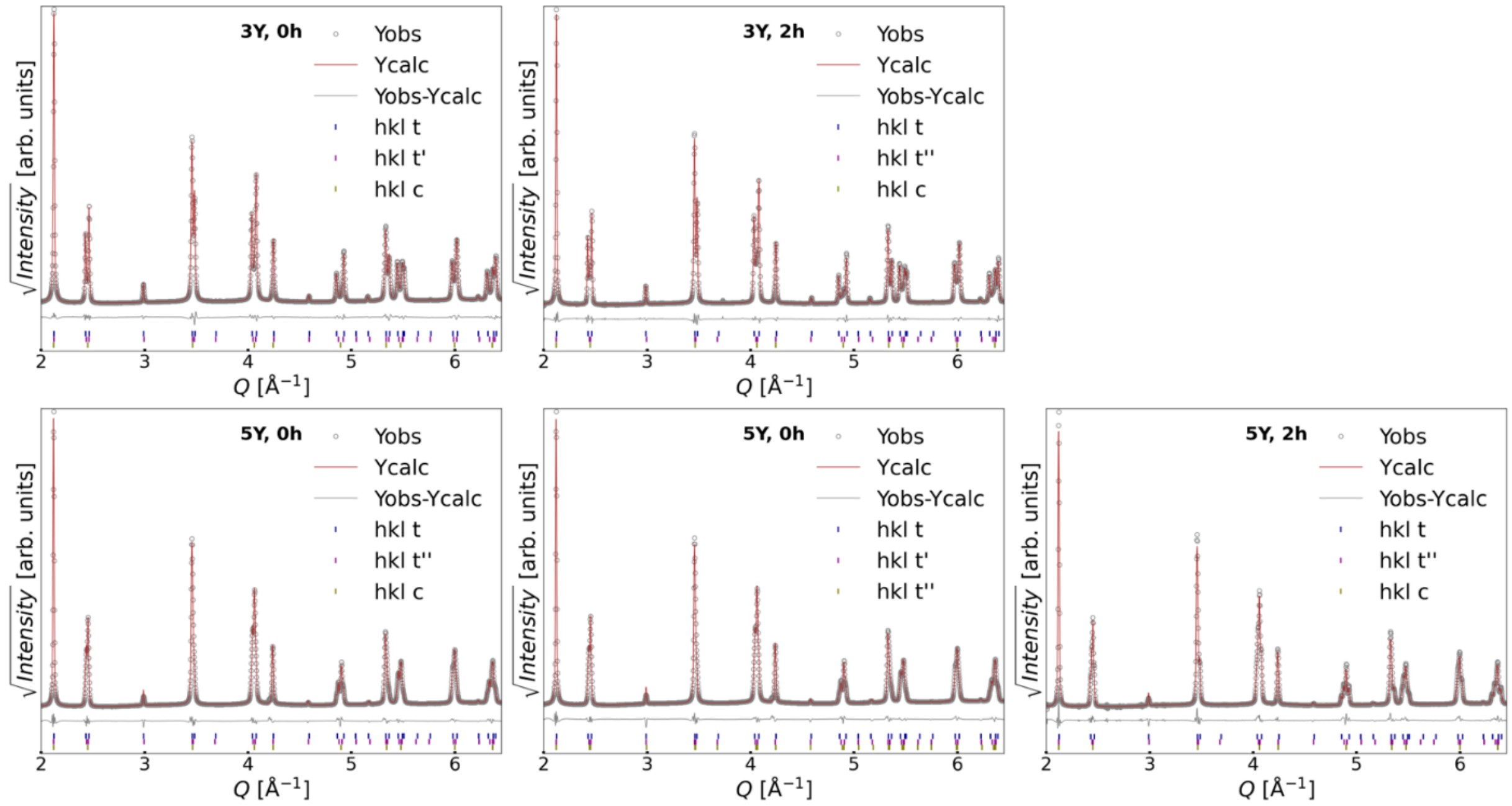

*Figure S6. XRD data with Rietveld refinement fits for regular the 3Y and 5Y samples presented in Figure 2. The sample and refined phases are indicated in the figure.*

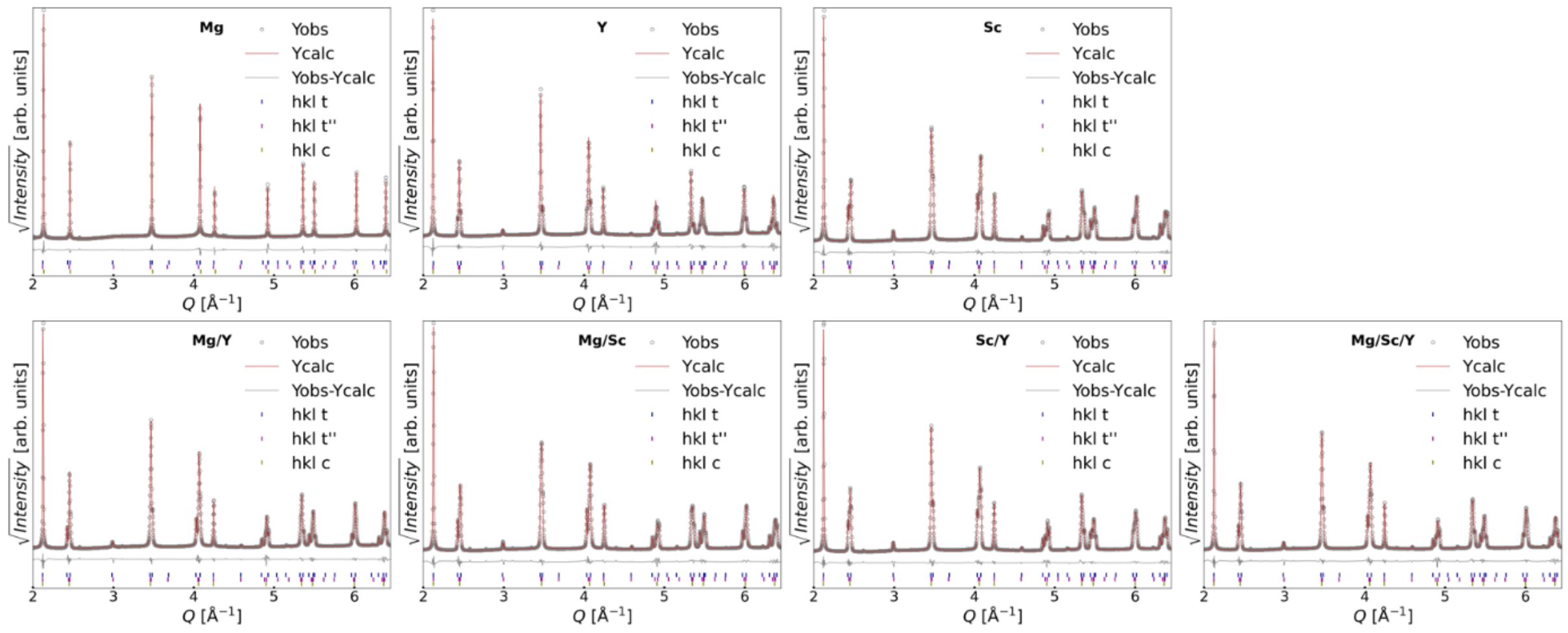

*Figure S7. XRD data with Rietveld refinement fits for infiltrated samples in Figure 3. The sample and refined phases are indicated in the figure.*

## S3.6 Input file for TOPAS

Below is the input-file for the 5Y-2h sample, as an example for all the samples refined with the same parameters.

```
'Input file for The Dip
r_wp  6.13516078 r_exp  0.494551251 r_p  4.67172322 weighted_Durbin_Watson  0.866973642 gof  12.4045501
iters 100000
chi2_convergence_criteria 0.00000001
do_errors

xdd 1_001_0001p_xye_new_250_lam3_fit_yresidual.xye '1_001_0001p_xye_new_250.xye
        x_calculation_step = Yobs_dx_at(Xo); convolution_step 1
        bkg @  35.60879`_0.132311966  1.02601776`_0.235592856 -6.72235446`_0.217499128  1.96877023`_0.208541972 -0.453004631`_0.200711622  0.558357546`_0.195625873  0.086352602`_0.190709358 -0.585610844`_0.189229437  0.857265709`_0.186197333  -1.36651194`_0.185679808  0.828466601`_0.185062841  -0.661853047`_0.186908108  0.0265759308`_0.186491341  -1.20597582`_0.187823484 0.975366884`_0.18506897  -1.2214541`_0.182073862  1.587056`_0.174703209  -0.440619884`_0.168837045  0.00131333515`_0.157492519  -1.5508585`_0.152940556  1.70548381`_0.141282225  -0.439803986`_0.136755227  0.0898459834`_0.114263782 -0.724048058`_0.108338735
        Simple_Axial_Model(!axial_Si_1, 1.34716845)
        lam ymin_on_ymax 0.0001
        la 1 lo 0.25509 lh 0.1 ' la = area, lo = lambda, lh = Lor HW
        LP_Factor(90)
        DSSD(u_1,-0.402787956`_0.00240370525, !s_1, 0)
        str
                scale scale_1t_250  6.9e-06`_6.967e-08_LIMIT_MAX_6.9e-06 max 6.90e-6
                weight_percent wp_1t  24.007`_0.271
                r_bragg  2.23415618
                Phase_Density_g_on_cm3( 6.04579242`_0.000307062979)
                TCHZ_Peak_Type(!pku_Si_250, 0.00913019334,!pkv_Si_250, -0.00883906731,!pkw_Si_250, -0.000719677473,!pkz_Si_250, 0.0000,!pkx_Si_250, 0.027389968, !pky_Si_250, 0.0019752965)
        str
                scale !scale_2t_250 1e-15' @  2.26315114e-06 min 5e-7
                weight_percent wp_2t          0.000`_0.000
                r_bragg  1.98457522
                Phase_Density_g_on_cm3( 6.00095734)
                TCHZ_Peak_Type(!pku_Si_250, 0.00913019334,!pkv_Si_250, -0.00883906731,!pkw_Si_250, -0.000719677473,!pkz_Si_250, 0.0000,!pkx_Si_250, 0.027389968, !pky_Si_250, 0.0019752965)
        str
                scale scale_3t_250  1.57547174e-05`_1.443e-07 'max 1.02e-5          '3.65984127e-06
                weight_percent wp_3t          54.526`_0.445
                r_bragg  1.90248303
                Phase_Density_g_on_cm3( 5.95643951`_0.000248197909)
                TCHZ_Peak_Type(!pku_Si_250, 0.00913019334,!pkv_Si_250, -0.00883906731,!pkw_Si_250, -0.000719677473,!pkz_Si_250, 0.0000,!pkx_Si_250, 0.027389968, !pky_Si_250, 0.0019752965)
        'Structural information from c:\Users\asbjorsf\OneDrive - Universitetet i Oslo\The dip\topas_vscode\cif_new_ASF_comeback\Y0_214Zr0_786O1_893_Fm-3m.cif
        str
                scale scale_c_250  1.55339506e-06`_4.772e-08 max 2.33e-6 '@  8.99994393e-07_LIMIT_MAX_9e-07 max 3e-7
                weight_percent wp_c          21.467`_0.531
                r_bragg  1.24561273
                Phase_Density_g_on_cm3( 5.95732725`_0.000479980855)
                TCHZ_Peak_Type(!pku_Si_250, 0.00913019334,!pkv_Si_250, -0.00883906731,!pkw_Si_250, -0.000719677473,!pkz_Si_250, 0.0000,!pkx_Si_250, 0.027389968, !pky_Si_250, 0.0019752965)

xdd 1_001_0001p_xye_new_700_lam3_fit_yresidual.xye '1_001_0001p_xye_new_700.xye
        x_calculation_step = Yobs_dx_at(Xo); convolution_step 4
        bkg   @     9.58334655`_0.0516596483     2.14574077`_0.0922682251   -0.582176125`_0.0872031304     0.190846458`_0.0822314362   -0.187392978`_0.0797693193   -0.168601202`_0.0759813165 0.31109083`_0.0754928597  -0.288308989`_0.0725682257  -0.191449882`_0.0752488931   0.0235816372`_0.0723014244   0.468629367`_0.0736885625   0.06167629`_0.0668276595   0.0746086423`_0.0652506649  -0.125339754`_0.0542345774  0.0937012968`_0.053768758
        Simple_Axial_Model(!axial_Si_2, 0.000100211345_LIMIT_MIN_0.0001)
        lam ymin_on_ymax 0.0001
        la 1 lo 0.25509 lh 0.1 ' la = area, lo = lambda, lh = Lor HW
        LP_Factor(90)
        DSSD(u_2,-0.272612074`_0.00233183435, !s_2, 0)
        str
                scale scale_1t_700  1.58586056e-06`_1.482e-08
                r_bragg  4.37043938
                Phase_Density_g_on_cm3( 6.04579242`_0.000307062979)
    TCHZ_Peak_Type(!pku_700, 0.00141591862,!pkv_700, -0.00204394997,!pkw_700, -0.000245701217,!pkz_700, 0.0000,!pkx_700, 0.0322430118, !pky_700, 0.000786172481)
        str
                'Known_Weight_Percent(wp_2t)
                scale scale_2t_700 = (wp_2t/(100-wp_2t)) Get(sum_smvs_minus_this) / (Get(cell_mass) Get(cell_volume));
                Phase_Density_g_on_cm3( 6.00095734)
                r_bragg  3.65289947
                TCHZ_Peak_Type(!pku_700, 0.00141591862,!pkv_700, -0.00204394997,!pkw_700, -0.000245701217,!pkz_700, 0.0000,!pkx_700, 0.0322430118, !pky_700, 0.000786172481)
        str
                'Known_Weight_Percent(wp_3t)
                scale scale_3t_700 = (wp_3t/(100-wp_3t)) Get(sum_smvs_minus_this) / (Get(cell_mass) Get(cell_volume));
```

```
                                        r_bragg  4.27971397
                                        Phase_Density_g_on_cm3( 5.95643951`_0.000248197909)
        TCHZ_Peak_Type(!pku_700, 0.00141591862,!pkv_700, -0.00204394997,!pkw_700, -0.000245701217,!pkz_700, 0.0000,!pkx_700, 0.0322430118, !pky_700, 0.000786172481)
                        str
                                        'Known_Weight_Percent(wp_c)
                                        scale scale_c_700 = (wp_c/(100-wp_c)) Get(sum_smvs_minus_this) / (Get(cell_mass) Get(cell_volume));
                                        r_bragg  3.61350649
                                        Phase_Density_g_on_cm3( 5.95732725`_0.000479980855)
                                        TCHZ_Peak_Type(!pku_700, 0.00141591862,!pkv_700, -0.00204394997,!pkw_700, -0.000245701217,!pkz_700, 0.0000,!pkx_700, 0.0322430118, !pky_700, 0.000786172481)

xdd 1_001_0001p_xye_new_970_lam3_fit_yresidual.xye '1_001_0001p_xye_new_970.xye
                        x_calculation_step = Yobs_dx_at(Xo); convolution_step 2
                        bkg    @      5.6322502`_0.0692812422      1.27536115`_0.132176088     -0.0940624901`_0.125946598      0.121860301`_0.117373451      0.097922602`_0.110418803    -0.424930289`_0.103653161
0.0323794765`_0.102158411  -0.0524822489`_0.0970409552    0.125584272`_0.0958948184  -0.182980225`_0.0895758227    0.0968011578`_0.0873430874  -0.0706177255`_0.0743336192    0.29025824`_0.0690059222  -
0.114310277`_0.0532817269 -0.0655362783`_0.0457157661
                        Simple_Axial_Model(!axial_Si_3, 0.0001_LIMIT_MIN_0.0001)
                        lam ymin_on_ymax 0.0001
                        la 1 lo 0.25509 lh 0.1 ' la = area, lo = lambda, lh = Lor HW
                        LP_Factor(90)
                        DSSD(u_3,-0.2456697`_0.00236860856, !s_3, 0)
                        str
                                        scale scale_1t_970  9.07782957e-07`_8.697e-09
                                        r_bragg  3.76886093
                                        Phase_Density_g_on_cm3( 6.04579242`_0.000307062979)
        TCHZ_Peak_Type(!pku_970, 0.0114937222,!pkv_970, -0.00272044109,!pkw_970, -0.000117204291,!pkz_970, 0.0000,!pkx_970, 0.0360441137, !pky_970, 0.000282514873_LIMIT_MIN_0.0001)
                        str
                                        'Known_Weight_Percent(wp_2t)
                                        scale scale_2t_970 = (wp_2t/(100-wp_2t)) Get(sum_smvs_minus_this) / (Get(cell_mass) Get(cell_volume));
                                        r_bragg  3.21307486
                                        Phase_Density_g_on_cm3( 6.00095734)
        TCHZ_Peak_Type(!pku_970, 0.0114937222,!pkv_970, -0.00272044109,!pkw_970, -0.000117204291,!pkz_970, 0.0000,!pkx_970, 0.0360441137, !pky_970, 0.000282514873_LIMIT_MIN_0.0001)
                        str
                                        'Known_Weight_Percent(wp_3t)
                                        scale scale_3t_970 = (wp_3t/(100-wp_3t)) Get(sum_smvs_minus_this) / (Get(cell_mass) Get(cell_volume));
                                        r_bragg  5.78079797
                                        Phase_Density_g_on_cm3( 5.95643951`_0.000248197909)
        TCHZ_Peak_Type(!pku_970, 0.0114937222,!pkv_970, -0.00272044109,!pkw_970, -0.000117204291,!pkz_970, 0.0000,!pkx_970, 0.0360441137, !pky_970, 0.000282514873_LIMIT_MIN_0.0001)
                        str
                                        'Known_Weight_Percent(wp_c)
                                        scale scale_c_970 = (wp_c/(100-wp_c)) Get(sum_smvs_minus_this) / (Get(cell_mass) Get(cell_volume));
                                        r_bragg  5.17145706
                                        Phase_Density_g_on_cm3( 5.95732725`_0.000479980855)
                                        TCHZ_Peak_Type(!pku_970,      0.0114937222,!pkv_970,      -0.00272044109,!pkw_970,      -0.000117204291,!pkz_970,      0.0000,!pkx_970,      0.0360441137,      !pky_970,
0.000282514873_LIMIT_MIN_0.0001)

for xdds {
for strs 1 to 1 {
phase_name t
                                        a   lpa_t  3.604175`_0.000080
                                        b   lpa_t  3.604175`_0.000080
                                        c  lpc_t  5.172856`_0.000128
                                        al 90.
                                        be 90.
                                        ga 90.
                                        volume  67.196`_0.003
                                        space_group "P42/nmcZ"
                                        site Zr1    x 0.75        y 0.25        z 0.75        occ Zr+4 0.951     beq  beq_Zrt  0.42475`_0.02183
                                        site Y1     x 0.75        y 0.25        z 0.75        occ Y+3  0.049     beq  beq_Zrt  0.42475`_0.02183
                                        site Sc1    x 0.75        y 0.25        z 0.75        occ Sc+3 0.000     beq  beq_Zrt  0.42475`_0.02183
                                        site Mg1    x 0.75        y 0.25        z 0.75        occ Mg+2 0.000     beq  beq_Zrt  0.42475`_0.02183
                                        site O1     x 0.25        y 0.25        z 0.4587      occ O-2  0.9755    beq  beq_Ot   1.09619`_0.08788
                                        LVol_FWHM_CS_G_L( 1, 4489.51014`_886765.188, 0.89, 5424.90744`_1389616.05,size_G_t, 10000`_4581057.88_LIMIT_MIN_0.3,size_L_t, 10000`_67573.0111_LIMIT_MIN_0.3)
                                        e0_from_Strain( 0.000722995575`_2.7841514e-05,strain_G_t, 0.303618452`_0.0086925023,strain_L_t, 0.0511463006`_0.0159522754)
}
for strs 2 to 2 {
phase_name t_prime
                                        a  !lpa_tt  3.611404
                                        b  !lpa_tt  3.611404
                                        c  !lpc_tt  5.155685
                                        al 90.
                                        be 90.
                                        ga 90.
                                        volume  67.242
                                        space_group "P42/nmcZ"
                                        site Zr1    x 0.75        y 0.25        z 0.75        occ Zr+4 0.906     beq !beq_Zrtt  0.94891 '1.04990
                                        site Y1     x 0.75        y 0.25        z 0.75        occ Y+3  0.094     beq !beq_Zrtt  0.94891 '1.04990
                                        site Sc1    x 0.75        y 0.25        z 0.75        occ Sc+3 0.000     beq !beq_Zrtt  0.94891 '1.04990
                                        site Mg1    x 0.75        y 0.25        z 0.75        occ Mg+2 0.000     beq !beq_Zrtt  0.94891 '1.04990
                                        site O1     x 0.25        y 0.25        z 0.4587      occ O-2  0.953     beq !beq_Ott   1.95576 '1.43834
                                        LVol_FWHM_CS_G_L( 1, 112.030012, 0.89, 106.69283, !size_G_tt, 120.713169_4005.26004, !size_L_tt, 10000_18657217.4)
                                        e0_from_Strain( 0.00119743711,!strain_G_tt, 0.548806712_2.89760801, !strain_L_tt, 0.0001_4.72222566)
}
for strs 3 to 3 {
phase_name t_double_prime
                                        a  lpa_ttt  3.621589`_0.000075
                                        b  lpa_ttt  3.621589`_0.000075
                                        c  !lpc_ttt  5.147913_0.000146
                                        al 90.
                                        be 90.
                                        ga 90.
                                        volume  67.520`_0.003
                                        space_group "P42/nmcZ"
                                        site Zr1    x 0.75        y 0.25        z 0.75        occ Zr+4 0.884     beq  beq_Zrttt  0.74950`_0.01565
                                        site Y1     x 0.75        y 0.25        z 0.75        occ Y+3  0.116     beq  beq_Zrttt  0.74950`_0.01565
                                        site Sc1    x 0.75        y 0.25        z 0.75        occ Sc+3 0.000     beq  beq_Zrttt  0.74950`_0.01565
                                        site Mg1    x 0.75        y 0.25        z 0.75        occ Mg+2 0.000     beq  beq_Zrttt  0.74950`_0.01565
                                        site O1     x 0.25        y 0.25        z 0.4587      occ O-2  0.942     beq  beq_Ottt   2.55510`_0.12139
                                        LVol_FWHM_CS_G_L( 1, 4489.51014`_723434.592, 0.89, 5424.90744`_1133613.95,size_G_ttt, 10000`_3737015.3,size_L_ttt, 10000`_64835.2655)
                                        e0_from_Strain( 0.000549033958`_2.0967645e-05,strain_G_ttt, 0.251600602`_0.00611145383,strain_L_ttt, 0.0001`_0.0127827805_LIMIT_MIN_0.0001)
}
for strs 4 to 4 {
phase_name c
                                        a  lpa_c  5.128767`_0.000138
                                        b  lpa_c  5.128767`_0.000138
                                        c  lpa_c  5.128767`_0.000138
                                        al 90.
                                        be 90.
                                        ga 90.
                                        volume  134.908`_0.011
                                        space_group "Fm-3m"
                                        site Zr1    x 0           y 0           z 0           occ Zr+4 0.786     beq beqZrc  3.06822`_0.09180
                                        site Y1     x 0           y 0           z 0           occ Y+3  0.214     beq beqZrc  3.06822`_0.09180
                                        site O1     x 0.25        y 0.25        z 0.25        occ O-2  0.946     beq beqOc   0.59403`_0.25070
```

```
        LVol_FWHM_CS_G_L( 1, 4489.51014`_2718369.72, 0.89, 5424.90744`_4259800.31,size_G_c, 10000`_14042911.6,size_L_c, 10000`_217587.624_LIMIT_MIN_0.3)
        e0_from_Strain( 0.00069528025`_7.30865255e-05,strain_G_c, 0.318634964`_0.0208789115,strain_L_c, 0.0001`_0.0451483552_LIMIT_MIN_0.0001)
}}
```

## S3.7 Y-content in the different phases

The quantity of $YO_{1.5}$ in each of the phases t, t'' and c where calculated following the equations from Krogstad et al. [32]. For the tetragonal phases

$$\frac{c}{\sqrt{2}a} = 1.02311 - 0.001498x$$

where c and a are the lattice parameters of the tetragonal unit cell, and *x* is the mol % $YO_{1.5}$ in the unit cell. For the cubic phase

$$a = 5.11742 + 0.001559x$$

where a is the cubic unit cell parameter and *x* is the mol % $YO_{1.5}$.

*Table S7. Calculated Y-content in the different phases for the different samples based on the equations provided by Krogstad et al. [32].*

| | Y-dip | 3Y 0h | 3Y 2h | 5Y 0h, t, t'', c | 5Y 2h | 5Y 0h, t, t', t'' |
|---|---|---|---|---|---|---|
| t | 5.2 | 5.5 | 4.9 | 6.8 | 5.5 | 6.8 |
| t' | | 9.1 | | | | 11.0 |
| t'' | 11.6 | | 11.9 | 11.3 | 12.0 | 12.9 |
| c | 10.6 | | 0.0 | 7.6 | 7.5 | 0.0 |
| total | 9.9 | 6.4 | 6.2 | 9.4 | 9.5 | 10.7 |

## S3.8 Volume model for the calculation of infiltration quantities

The pseudo cubic cell volume of the fluorite cation framework follows from the cation anion touching condition along the body diagonal. The cubic cell parameter equals four times the sum of the average cation radius and the oxide ion radius, divided by the square root of three. The cell volume is therefore

$$V = \left(\frac{4}{\sqrt{3}}\right)^3 \cdot (\bar{r}_{cation} + r_O)^3 \text{ (Eq. 1)}$$

Here $r_O$ is the Shannon four coordinate oxide ion radius of 1.38 Å, and $\bar{r}_{cation}$ is the Vegard mean of the cation radii. The mean cation radius is

$$\bar{r}_{cation} = x_{Zr}\, r_{Zr} + x_Y\, r_Y + x_{Sc}\, r_{Sc} + x_{Mg}\, r_{Mg} \qquad \text{(Eq. 2)}$$

where $x_i$ is the cation fraction of species i and $x_{Zr}$ equals one minus the sum of the dopant cation fractions. The model is purely geometric and does not include an explicit oxygen vacancy contraction term. The vacancy effect is folded into the chosen value of each cation radius through the effective coordination number assigned to that cation.

### Ionic radii

The model uses four cation radii and one anion radius. Zirconium is the Shannon eight coordinate value of 0.84 Å [58]. Yttrium is set to 0.95 Å, derived from the EXAFS Y to O bond length of 2.33 Å in tetragonal and cubic yttria stabilized zirconia reported by Li, Chen and Penner-Hahn [26–28]. Yttrium remains eight coordinated in this matrix because oxygen vacancies cluster around Zr rather than Y. Scandium is set to 0.81 Å, the Shannon seven coordinate interpolation between CN 6 and CN 8. The value is supported by the bimodal CN 6 and CN 8 distribution observed in 8 mol % scandia stabilized zirconia by 45Sc magic angle spinning NMR by Kim, Hsieh and Stebbins, where the population average gives an effective seven coordinate environment [57]. Magnesium is set to 0.78 Å, the midpoint of the physically reasonable window 0.72 to 0.81 Å. Three arguments fix this window. First, $Mg^{2+}$ has a strong preference for octahedral coordination, shown by the rock salt structure of MgO with Mg O bond length 2.105 Å, and the corresponding Shannon CN 6 radius is 0.72 Å [58]. Second, $Mg^{2+}$ on the $Zr^{4+}$ site carries an effective charge of minus two, twice that of $Y^{3+}$ or $Sc^{3+}$, so the Coulomb attraction between Mg and the compensating oxygen vacancy of effective charge plus two is the strongest of the three dopants. This favours close pairing of Mg with a vacancy and supports a near sixfold local environment. Third, charge balance requires one oxygen vacancy per $Mg^{2+}$, twice the vacancy yield of $Y^{3+}$ or $Sc^{3+}$, which reinforces the same picture. The lower bound 0.72 Å is the rigid CN 6 value. The upper bound 0.81 Å is the Shannon CN 7 interpolation. The midpoint 0.78 Å is taken as a single representative value. Oxygen is taken at the Shannon four coordinate value of 1.38 Å [58].

## Composition partition

The Y-dopant fraction per phase is set by the Krogstad [32] relation for tetragonality and Y-content

$$\frac{c}{a \cdot \sqrt{2}} = 1.02311 - 0.1498\, x_{eq} \qquad \textit{(Eq. 3)}$$

## Calculation steps

The measured tetragonality of each phase is taken into Eq. 3 to give $x_{eq}$, which is only valid for Y as dopant.

The calculated volume, $V_{calc}$ from Eq. 1., of each phase is calculated based on input radii $\bar{r}_{cation}$ determined by the dopants present and their quantities. Due to the inaccuracy of the model, the calculated volume of the t-phase, $V^t_{calc}$, is adjusted to fit the with the observed volume, $V^t_{obs}$. The adjusted volume is given by

$$V_{adjusted} = V_{calc} - (V^t_{calc} - V^t_{obs}) \qquad \textit{(Eq. 4)}$$

Which for the t-phase imply that $V_{adjusted} = V^t_{obs}$.

We then considering the difference in volume between the t-phase ($V^t_{calc}$) and the other phases so that the goal of the model is to obtain

$$\frac{V^{phase}_{calc}}{V^t_{calc}} = \frac{V^{phase}_{obs}}{V^t_{obs}} \qquad \textit{(Eq. 5)}$$

Where $V^{phase}_{calc}$ and $V^{phase}_{obs}$ are the calculated and observed volumes of the phases *t′*, *t′′* and *c*. The model minimizes the difference of this equation to obtain suitable volumes.

The volumes $V_{calc}$ are calculated by Eq. 1 and Eq. 2, and the concentration of Y, Mg and Sc are obtained. To make the model reasonable, certain restraints or constraints need to be applied.

One simplification is that the *t*-phase only contains Y and that the Y-content in this phase is determined by the Krogstad relation Eq. 3. Next, that the infiltrated elements are present in equal quantities, *i.e.* that $c_{Sc} = c_{Mg} = c_Y$ in the case where all three elements are present. Y present in the sample from 3YSZ precursor material is included as well. In this case, the difference between the volumes (Eq. 5) can be refined separately for the *t′*, *t′′* and *c*-phases, and a good fit to the experimental volumes can be obtained. However, the Y-content in the infiltrated samples are very hard to probe here as the large Y-cation with compensate for the smaller Sc and Mg cations, and there will be several solutions available. One option is to assume 5.4 % Y in all the phases, regardless of how correct it is. Clearly, the info obtained is not quantitative due to missing constraints for physical fact that we should know, such as the total dopant concentration, which we handle below.

Another possible constraint is the phase fractions. We can say that the Y-content calculated from the volume model, combined with the phase fractions, must sum to the quantity of Y we know are present in the sample. For the 3Y and 5Y samples, value is $c_{Y\ in\ 3Y} = 5.6\ \%$ and $c_{Y\ in\ 5Y} = 10.0\ \%$.

$$c_{Y\ in\ sample} = pf_t * x_Y^t + pf_{t'} * x_Y^{t'} + pf_{t''} * x_Y^{t''} + pf_c * x_Y^c \qquad \textit{(Eq. 6)}$$

Where $pf$ is the phase fraction. In this case, one minimization is performed where all variables are fitted to one set of constraints. However, this restricts the model so much that the results might not be more informative. The equation above can be expanded to include the Sc and Mg based on the measured infiltration quantities. However, the data does not manage to reproduce the shape of the volume curves, and such quantitative detail is impossible to extract. This is both due to data quality and poor local structural model to correctly represent the volume effects of introducing various dopants. Values obtained for both models for Y-only containing samples are presented in Table S8.

*Table S8. Values from volume model.*

| Sample | Phase | x(Y) no restriction | phase fraction | x(Y) *pf | x(Y) restricted |
|---|---|---|---|---|---|
| **3YSZ-0h** | t | 0.05527 | 0.762 | 0.04156 | 0.05454 |
| **3YSZ-0h** | t' | 0.06142 | 0.238 | 0.01444 | 0.06069 |
| **3YSZ-2h** | t | 0.04873 | 0.814 | 0.04091 | 0.05025 |
| **3YSZ-2h** | t'' | 0.07963 | 0.186 | 0.01509 | 0.08115 |
| **5YSZ-0h(t,t',t'')** | t | 0.06796 | 0.244 | 0.02003 | 0.08208 |
| **5YSZ-0h(t,t',t'')** | t' | 0.09218 | 0.412 | 0.04522 | 0.10976 |
| **5YSZ-0h(t,t',t'')** | t'' | 0.08688 | 0.344 | 0.03475 | 0.10102 |
| **5YSZ-0h(t,t'',c)** | t | 0.06762 | 0.266 | 0.02048 | 0.07701 |
| **5YSZ-0h(t,t'',c)** | t'' | 0.08946 | 0.544 | 0.05846 | 0.10746 |
| **5YSZ-0h(t,t'',c)** | c | 0.08857 | 0.215 | 0.02106 | 0.09794 |
| **5YSZ-2h** | t | 0.05521 | 0.242 | 0.01715 | 0.07085 |
| **5YSZ-2h** | t'' | 0.08703 | 0.553 | 0.06291 | 0.11376 |
| **5YSZ-2h** | c | 0.08169 | 0.205 | 0.01995 | 0.0973 |
| **Y-inf** | t | 0.05234 | 0.162 | 0.00848 | 0.05234 |
| **Y-inf** | t'' | 0.07932 | 0.233 | 0.01848 | 0.07932 |
| **Y-inf** | c | 0.08586 | 0.605 | 0.05194 | 0.08586 |

## Suggested results table

A single table with one row per sample and phase keeps the layout compact and lets the reader compare the model and the data along the same axis. Proposed columns are sample identifier, phase label, tetragonality, observed volume, $x_{eq}$ from Eq. 3, the cation fractions $x_Y$, $x_{Sc}$ and $x_{Mg}$, the mean cation radius $\bar{r}_{cation}$, the calculated volume from Eq. 1, and the residual $V_{obs}$ minus $V_{calc}$. An example is given below.

| Sample | Phase | tet | V_obs (Å³) | x_eq | x_Y | x_Sc | x_Mg | r̄_cation (Å) | V_calc (Å³) | V_obs − V_calc (Å³) |
|---|---|---|---|---|---|---|---|---|---|---|
| 3YSZ-0h | t | 1.01483 | 134.36 | 0.0553 | 0.0553 | 0.000 | 0.000 | 0.846 | 135.87 | −1.51 |
| Sc-inf | c | 1.0000 | 134.84 | 0.1543 | 0.0560 | 0.0478 | 0.000 | 0.845 | 135.62 | −0.78 |
| Mg/Sc-inf | c | 1.0000 | 134.41 | 0.1543 | 0.0560 | 0.0397 | 0.0397 | 0.843 | 135.24 | −0.83 |

$V_{calc}$ above is the unanchored prediction from Eq. 1. The residual column shows $V_{obs}$ minus $V_{calc}$. An anchored version replaces $V_{calc}$ by the calculated volume shifted by the t-phase residual of the same sample, in which case the residual reports the relative model error between phases of the same sample.